\documentclass[11pt,fleqn]{article}

\date{}

\begin{document}

\title{Towards gravitating discs around stationary black holes}
\author{{\sc Old\v{r}ich Semer\'ak} \vspace{3mm} \\
 {\it Institute of Theoretical Physics,
      Faculty of Mathematics and Physics,} \\
 {\it Charles University, V Hole\v{s}ovi\v{c}k\'ach 2,
      CZ-180 00 Praha 8, Czech Republic} \\
 {\it E-mail: oldrich.semerak@mff.cuni.cz}}
\maketitle

\begin{abstract}
This article outlines the search for an exact general
relativistic description of the exterior (vacuum)
gravitational field of a rotating spheroidal black hole
surrounded by a realistic axially symmetric disc of matter.
The problem of multi-body stationary spacetimes is first exposed
from the perspective of the relativity theory (section
\ref{multiple}) and astrophysics (section \ref{astrophysical}),
listing the basic methods employed and results obtained.
Then (in section \ref{formulas}) basic formulas for stationary
axisymmetric solutions are summarized.
Sections \ref{static} and \ref{stationary} review what we have
learnt with Miroslav \v{Z}\'a\v{c}ek and Tom\'a\v{s} Zellerin
about certain static and stationary situations recently.
Concluding remarks are given in section \ref{concluding}.
Although the survey part is quite general, the list of references
cannot be complete. Our main desideratum was the informative
value rather than originality --- novelties have been preferred,
mainly reviews and those with detailed introductions.
\end{abstract}

\section{The fields of multi-body systems involving black holes}
\label{multiple}

The subject of self-gravitating sources around rotating black
holes is interesting in several respects, relevant from the point
of view of the relativity theory itself as well as in the
astrophysical context.

First, due to the non-linearity of Einstein's equations, the
field of a multi-body system is a traditional challenge where one
typically does not manage with a simple superposition.\footnote
{Even the problem of an ``isolated'' object is very difficult
 (mainly if its motion must be found as well, not mentioning
 the interior field) unless one makes some simple assumptions
 about its multipole structure;
 see e.g. \cite{Dix79,EhlR77,BaiI80} for a treatment of bodies
 with matter interior, \cite{ThoH85} for that also valid for
 singular bodies such as black holes, and \cite{BlaF01} for the
 case of point-like particles.}
On the first post-Newtonian level, the ``celestial mechanics'' of
gravitationally interacting bodies can be kept linear
\cite{DamSX91,DamSX92}, but in the strong-field region the
interaction may bring surprising features that have only been
described in a very few cases yet (for a two-body problem ---
today at the centre of attention because of the expected
gravitational waves from colliding compact binaries, ``the
current state of art is the third post-Newtonian approximation''
\cite{AndBF01}).

The second point in which the problem embodies the essence of
general relativity is the effect of inertial frame dragging due
to the rotation of the sources. Contrary to the Newtonian
treatment which does not discriminate between static and
stationary situation, the field is now determined not only by
mass-energy configuration, but also by its motion within the
bodies. The inertial space can be imagined as a viscous fluid
mixed by the sources. In today's ``gravito-electromagnetic''
language, gravity has not only an electric component, generated
by the mass, but also a magnetic one, generated by mass currents;
see
\cite{Pen83,Emb84,Mas85,ThoPM86,JanCB92,CiuW95,LynKB95,LynBK99,
KinP01}
and references therein. (The long-time effort to measure the
gravitomagnetic effect of the Earth is just culminating
\cite{Ciu00}; cf. \cite{Nor88,Nor99}.)

The third point is that our setting involves a black hole,
perhaps the most bizarre of the new predictions of general
relativity.\footnote
{The prologue to a major reference \cite{Chan83} (not speaking
 about its entire content) however convinces us that black holes
 are ``the simplest objects'' (also ``the most perfect
 macroscopic objects'') in the universe.
 Cf. also \cite{ThoPM86,FroN98}.}
Within and around it, the deviations from Newtonian theory become
dominant, in particular the above mentioned implications of
non-linearity and rotation reveal themselves prominently.
Perhaps the most extreme example (of an exact spacetime)
involving {\em all the three} aspects\footnote
{Within general relativity, there is only one more ingredient
 that could make the situation even more dramatic:
 non-stationarity. The result would be a gravitational collapse
 or a collision of already collapsed objects. These events are
 currently devoted an exceptional attention as the strongest
 sources of gravitational waves.}
is a ``double Kerr(-Newman)'' solution for an aligned pair of
rotating ultracompact objects \cite{DieH85,ManR01,BicH85,ManMR94}
(cf. \cite{KriP98} where an approximate method for axially
symmetric superpositions of rotating black holes was proposed as
a starting point of computations of black hole collisions).

Relativistic spacetimes are found in three ways:
by numerical solution of Einstein equations,
by perturbation of previously known spacetimes, and
by exact analytical solution of Einstein equations.
Let us mention what the above approaches told us about stationary
axisymmetric spacetimes describing rotating black holes with
additional matter (ring, disc or torus).\footnote
{In the following, we do not review superpositions, obtained by
 any of the methods, of black holes with external
 {\em electromagnetic} fields.
 We refer to \cite{BicD77,BicD76,Lin79,BicD80} for test EM
 fields and to \cite{EstR88,AliG89,AleG96} for exact solutions.
 A more recent account can be found in \cite{BicL00}.}

\subsection{Numerical solution}

Although computers have been naturally employed in relativity for
decades, in algebra as well as in numerics, it is only quite
recently that they embarked on numerical solution of Einstein's
equations in the most interesting cases involving black holes.
Teukolsky \cite{Teu98} lists today's peak parallel-type hardware
and software and the appropriate, hyperbolic formulation of the
field equations as ``three reasons why we are on the verge of
important advances in the computer solution of Einstein's
equations''. Whereas analytical prospects are restricted, the
present-day computational facilities can handle almost any
situation, at least for certain period of time. On the other
hand, it is never sure whether a given numerical solution
represents a typical or just a marginal case. Due to this lack of
generality, it is difficult to discern, analyse and interpret
different classes of solutions within boundless ranges of
possibilities. However, numerical solutions can provide explicit
{\em examples} of spacetimes and processes that might otherwise
remain only conjectural \cite{Leh01}.

Numerical spacetimes containing a rotating black hole surrounded
by an additional stationary axisymmetric source were constructed
by \cite{Lan92} (a hole with a thin finite equatorial disc) and
by \cite{NisE94} (a hole with a thick toroid). To mention just
one particular point, \cite{Lan92} ended with a prolate horizon
(stretching along the axis) in certain cases (when the disc was
strongly counter-rotating with respect to the hole), which was
not observed in \cite{NisE94}. The horizons have been known to
inflate towards the external sources, so it would be an
interesting consequence of the interplay between dragging from
the hole and from the disc if it were confirmed that they can
indeed become prolate, even though under extreme circumstances
only.

Note, however, that the main stream of numerical relativity is
focused on {\em non}-stationary problems important in generation
of gravitational waves, in particular on black hole inspiral
collision (e.g. \cite{Tho98,Teu98,AlcB01} and a thorough review
\cite{Leh01}).

\subsection{Perturbative solution}

A great deal of literature was devoted to perturbations of
black-hole spacetimes and several formalisms were developed, none
of which can be reviewed here.\footnote
{In monograph \cite{Chan83}, the author admits that ``\dots the
 account, in large parts, is hardly more than an outline'' at
 the beginning of chapter {\it The gravitational perturbations of
 the Kerr black hole} which has 101 pages.}
In \cite{Wil74}, a fully explicit solution for the perturbation
of the Schwarzschild metric by a (rotating) axisymmetric weakly
gravitating thin equatorial ring was found by solving the
perturbed Einstein equations. This direct approach is however not
practicable for a rotating hole and/or for an extended external
source with pressure (not mentioning more complicated
situations).

The rotating case (specifically, the algebraically special vacuum
case) was seized notably by Teukolsky \cite{Teu73} who succeeded
in separating the decoupled equations for the first order
perturbations of a Kerr black hole into the second-order ordinary
differential equations for the scalar field (for scalar
perturbations) or for the Newman-Penrose scalars constructed from
the electromagnetic field tensor (for electromagnetic
perturbations), from the neutrino spinor (for neutrino
perturbations) or from the Weyl tensor (for gravitational
perturbations). By solving the ``gravitational'' Teukolsky
equation, the perturbative deformation of the Kerr horizon was
calculated by \cite{Dem76}. The stationary axisymmetric Green's
function of the Teukolsky equation was provided by \cite{Lin77}.
The most comprehensive expositions of the first-order black hole
perturbation theory were given by Chandrasekhar
\cite{Chan79,Chan83} in Newman-Penrose formalism. More recently,
\cite{Lea86,ManST96,ManT97} solved the Teukolsky equation in
terms of Coulomb wave functions and hypergeometric functions.

In a related method of handling the first-order perturbations of
rotating fields, first devised for electromagnetic perturbations
by Cohen \& Kegeles and then generalized to gravitational ones
\cite{Chrz75}, the perturbation components are expressed in terms
of a single (Debye) potential that obeys a wavelike equation.
Wald showed that this equation is just the adjoint of the
Teukolsky master equation, while \cite{KegC79} gave a covariant
extension of the method of Debye superpotentials to neutrino,
electromagnetic and gravitational perturbations of all
algebraically special spacetimes (see \cite{Tor90,TorS99} for
summary, references and generalization). In the meantime,
\cite{Chrz76} treated the problem of perturbation of the Kerr
horizon after having learnt \cite{Chrz75} how to fix the
perturbed metric from the Newman-Penrose functions. General
perturbations of Schwarzschild solution were treated in this
manner by \cite{Tor96}. Latest advance of the approach consists
mainly in formulating the conservation laws for perturbations
\cite{TorS99,Car00}.

Let us mention yet another approach \cite{SarHB01}, also dealing
with curvature components but not restricted to algebraically
special backgrounds. It has been shown, in the static case and to
the linear order so far, that it can be generalized naturally to
self-gravitating matter fields; in a spherically symmetric case,
the connection with older, metric formalisms of Regge \& Wheeler
and Zerilli was elucidated.

In the charged case, the perturbation problem introduces coupling
between gravitational and electromagnetic quantities, implying
e.g. conversion between gravitational and electromagnetic waves.
Interacting perturbations of the Reissner-Nordstr\"om black hole
were studied by \cite{Bic79} (using the earlier formalisms of
Regge \& Wheeler, Zerilli and Moncrief) and by \cite{TorC96}
(using the method of scalar potentials). A gauge invariant
derivation of the basic equations of different formalisms was
provided in \cite{FerL96}, while \cite{FerL97} related their
rotating analog to Teukolsky equation.

The second-order perturbation theory has also been under
development since 1970s. For Schwarzschild black holes it is
surveyed in \cite{GleNPP00}, while the rotating case is tackled
in \cite{CamL99}.

It should be remarked that solutions describing stationary
sources around black holes were not the primary aim of the
black-hole perturbation strategy. Historically, attention has
rather been devoted to the stability of black-hole solutions
\cite{PreT73}, to the relaxation of black holes into a stationary
``no hair'' state \cite{Hod00}, involving the issue of back
reaction of a perturbation \cite{AbrF01} propagating out of
\cite{Bar99} or into the hole \cite{Ori97,Ori99} (cf.
\cite{Isr98}), to non-stationary processes of astrophysical
significance such as the behaviour of weakly gravitating
particles or waves in black-hole backgrounds \cite{Poi97,Jez99},
and, of course, to gravitational waves \cite{NakOK87} ---
presently mainly to black-hole collisions as their most prominent
source (e.g. \cite{CamKLPR01,CamGHWZ01,Kha01}; for a survey, see
\cite{Tho98}). The latter is exactly an example of a problem
where numerical relativity and perturbative analysis may --- and
should --- coexist while mutually checking each other
\cite{Sei99,Baketal00,Lou01}.

\subsection{Exact solution}

Perturbative approach is adequate in situations when the external
source has only a very small effect on the field of the main
body. Whenever this is not the case, the superpositions have to
be described by solutions of the full, non-linear theory.
At the end of Ji\v{r}\'{\i} Bi\v{c}\'ak's recent survey
\cite{Bic00}, one is encouraged ``not to cease in embarking upon
journeys for finding them, and perhaps even more importantly, for
revealing new roles of solutions already known'', because ``Is
there another so explicit way of how to learn more about the rich
possibilities embodied in Einstein's field equations?'' For a
more technical demonstration, see the canonical monograph
\cite{KraSHM80} (an updated version is awaited); more recent
reviews of vacuum fields can be found in \cite{Bon92,BonGM94}.

The only stationary (in fact static) equilibrium configuration
containing more than one black hole is the Majumdar-Papapetrou
case with extreme centres of the Reissner-Nordstr\"om type (e.g.
\cite{Gib80,PerC97,BreMS98,Chru99}), where the electrostatic
repulsion exactly compensates the gravitational attraction. With
rotating sources, gravitomagnetic interaction between spins
(repulsive in the parallel case --- see e.g. \cite{PfiS87}) and
magnetic interaction between magnetic dipole moments (repulsive
in the antiparallel case) are also present. Analysis of the
momentarily stationary and axisymmetric system of two identical
sources was carried out by \cite{DieH85,ManR01} for the Kerr
components (with mass $M$ and specific rotational angular
momentum $a$) and by \cite{BicH85,ManMR94} for the Kerr-Newman
components (with $M$, $a$ and charge $Q$). Equilibrium was found
to be only possible for the extreme values of the charges
($Q=M$). In all other cases, supporting singularities
(``struts'') occur indicating that a given system of sources
cannot remain stationary (or static) according to the field
equations. Hence, rotating centres can only remain in equilibrium
in a super-extreme case of two naked singularities (that have
$Q^{2}+a^{2}>M^{2}$) \cite{BreM95}, both the magnetic
dipole-dipole and the gravitational spin-spin interactions being
too weak to keep apart black holes (having
$Q^{2}+a^{2}\leq M^{2}$).
(On the other hand, a non-singular equilibrium {\em is} possible
with two spinning Curzon particles \cite{DieH82}. Most recent
discussions of both cases, in particular of the character of
supporting singularities, can be found in
\cite{LetO98,Bon01a,Bon01b}.)

Accretion discs of astrophysical interest are likely to lie in
the equatorial plane of the centre and unlikely to bear a
considerable charge, so that neither the spin-spin nor the
electromagnetic repulsion can act within them. One must rather
refer to a centrifugal force resulting from the orbital motion of
the material or to hoop stresses when interpreting the situation.

It turned out to be difficult to superpose a Kerr centre with an
additional axisymmetric ring, disc or torus and no explicit exact
solutions describing the systems of this kind have been found
until now. Nevertheless, many stationary axisymmetric solutions
of the (electro)vacuum Einstein equations are known that do
generalize those containing only isolated black holes. These were
mostly obtained by indirect methods known as ``generating
techniques''. Two major approaches, developed by the end of the
1970's --- the group-theoretic techniques and the
soliton-theoretic (or inverse-scattering) techniques, work for
spacetimes with two commuting symmetries.
(See \cite{Kor99} for a review and
\cite{Cos80,Cos81,Cos82,HoeD84,Let89,Mai00} for more
detailed analysis and interrelations between different
formulations, e.g. those of Kinnersley \& Chitre; Maison;
Belinskii \& Zakharov; Harrison; Hoenselaers, Kinnersley \&
Xanthopoulos; Hauser \& Ernst; Neugebauer; Kramer \& Neugebauer;
or Alekseev.)
Other related methods have been proposed more recently, e.g.
the simplification of the Hauser-Ernst integral equation by
Sibgatullin \cite{Sib84}, ``monodromy data transform'' by
Alekseev \cite{Ale01} (also \cite{KleR98}), the static
gravitational multipoles \cite{GutM85} (cf. \cite{Que92}) and
their superposition with stationary fields \cite{GutM88} by
Gutsunaev \& Manko, the linear transformation by Quevedo
\cite{Que92} or the ``finite-gap'' (algebraic-geometric)
solutions by Korotkin \& Matveev (\cite{Kor97} and references
therein). Most of them are briefly compared in (appendix 6. of)
\cite{AleG96}.
(For other approaches, see e.g. \cite{Tan79,Cle00} or the results
of Nakamura and Kyriakopoulos, referred to and worked out by
\cite{Ter90}; cf. also \cite{Vei85}.)

The crucial point of the soliton generating techniques is a
solution of two linear differential equations (Lax pair) the
integrability conditions of which are exactly the Einstein's
equations (namely the Ernst equation). The linear problem can be
tackled in order to generate new solutions from the known ones:
given some (``seed'') metric with two Killing vectors, it yields
a different metric of the given type (the procedure is often
called B\"acklund transformation in analogy with the technique
used for the KdV equation). In such a manner, many known
spacetimes were reproduced, but also broad families of new
solutions were provided characterized by arbitrarily large sets
of free parameters. However, only a very restricted number of
them have been given a clear physical interpretation. Though
several of these results perhaps represent a rotating black hole
in an ``external'' gravitational field
(e.g. \cite{QueM91,Man92,ManN92,BreGMD98,ChauD97a,ChauD97b}),
none of the latter has yet been specified to the case generated
by a concrete realistic body such as ring, disc or torus (cf.
the last paragraph of section IV in \cite{LetO87}).

Making the ``random'' B\"acklund transformation with some metric,
it is, however, difficult to require specific physical properties
of the spacetime being constructed --- one rather ``takes pot
luck'' and looks what comes out (see \cite{Que92} for a hint how
to overcome this); section \ref{soliton} will illustrate what
problems may arise. It would be more appropriate to express the
physical boundary conditions of the Ernst equation in terms of
the quantities which appear in the linear problem and then solve
the latter. This leads to Riemann-Hilbert problems known from the
theory of completely integrable differential equations
\cite{Neu96,KleR98,AnsKMN02}. Tackling the stationary
axisymmetric boundary value problem with the help of the linear
system, the fields of two physically relevant types of sources
have been discussed: that of black holes \cite{Mei00,Neu00} and
that of finite thin discs (\cite{Mei00,Neu00,Ans01,FraK01} and
references therein). There is some hope that the above methods
could also be used to describe superpositions of both (e.g.
\cite{Neu00}).

\section{Astrophysical relevance}
\label{astrophysical}

The system of a black hole with an accretion disc is very
important in astrophysical considerations. In particular, it
plays the key role in models of a whole range of active galactic
nuclei and of some X-ray binaries.

\subsection{Galactic nuclei and X-ray binaries}
\label{AGN}

Extraordinary luminosity generated within a small volume (as
manifested by its rapid variability), the presence of very hard
(X and gamma) radiation, high rotation speeds and broad velocity
distribution of the material around indicate that there is an
interacting ultracompact object in the inner regions of these
sources \cite{Ree98,CelMS99,Lao99,Chak99}.\footnote
{In fact, the best evidence for such an object was provided by
 the nuclei of several ``normal'' galaxies \cite{KorG01} (led
 by our own \cite{GheMBTK00}) which are not so over-exposed with
 matter and radiation and where it is easier to study the motion
 of individual sources.}
The ambient gas, present in the galactic nucleus or overflowing
from the double-star primary, typically has enough angular
momentum not to fall directly (almost radially) onto the object;
it rather forms a disc and only gradually spirals down. The flow
is strongly sheared due to a considerable non-homogeneity of the
field, which enables viscous torques to heat the gas to high
temperatures.
(See e.g. \cite{Chak96,KatFM98,AbrBP98,AgoK00,Rif00} for various
aspects of the modern accretion disc theory. In particular, the
most realistic grid of non-LTE disc models was constructed in the
paper series \cite{HubH97,HubH98,HubABK00,HubBKA01}.)
In some cases, the excess of angular momentum drives outflows
through the empty regions along the rotation axis of the system.
Jets of matter-energy emanate from a number of active nuclei,
often at relativistic speeds, feeding giant outer lobes
distinguished in radio surveys. The formation mechanism of the
jets is studied by several groups using relativistic-MHD
numerical simulations (e.g. \cite{KoiMSK00,Bog01}).\footnote
{For the jet in M87, the first known and the best resolved one,
 simulations are already fitted to observations at distances from
 several dozens of Schwarzschild radii from the centre
 \cite{JunBL99,FenM01}!}
The high degree of jet collimation evidences an important role of
magnetic fields \cite{KarD97,Bog00,FelZ00,MeiKU01} and poses
questions concerning the interaction of jets with the ambient gas
and radiation.
(See chapter 9 of \cite{WheM01} and \cite{Liv99} for a review.)
The entire topic of active galactic nuclei is covered in
\cite{Kro99} (for earlier references, see \cite{Wii85})
while that of X-ray binaries in \cite{LewPH95}.
Recent status of both areas can be learned from \cite{KapHW01}.

In spite of the observational variety of active galaxies
(blazars, quasars, Seyfert galaxies, radiogalaxies, \dots), it
now seems that we simply look at a similar type of source from
different directions \cite{Elv00}. The discovery of jet
structures accompanying several X-ray binaries in our Galaxy (the
literature speaks of ``microquasars'' \cite{MirR99,Sin01}) has
strengthened the conviction that the stellar-size active sources
are also powered by the same accretion mechanism. Further
evidence is provided by observations of compact X-ray sources in
several nearby galaxies \cite{Maketal00}.

The suspicion of the presence of ultracompact objects has also
been supported by more detailed considerations. Certain spectral
features (the iron K$\alpha$ line in particular) of some active
galactic nuclei (a celebrated example is the Seyfert 1 galaxy
MCG--6-30-15) have been interpreted as originating as close to
the very centre as a few to a few dozens of Schwarzschild
radii,\footnote
{Even a distance of only half the Schwarzschild radius has
 already been announced (!), giving support for a rotating black
 hole.}
so they may well represent data from the strongest fields ever
met (see \cite{FabIRY00} for a review, and e.g.
\cite{Rus00,RusFRI00,KarCAA00,DabL01} for models and
interpretation issues). The iron lines have already been
discovered in the X-ray spectrum of several Galactic microquasars
(\cite{Miletal01a} and references therein).

Also, the frequencies of quasi-periodic oscillations, found at a
growing number of stellar-size sources, have been identified as
the values characteristic for the accretion flow close to a black
hole or a neutron star
\cite{KaaFCh97,MorRG97,PsaBK99,MatP99,CuiSHH00,Str01a,Str01b,
Miletal01b}.
This allows us to consider models involving orbital frequencies
of ``hot spots'' in the inner parts of an accretion disc
(e.g. \cite{Kar99}),
beat of these frequencies with the centre's spin
\cite{Psaetal99},
blobs on tilted orbits affected by Lense-Thirring precession
\cite{CuiZCh98,ArmN99,MerVSB99},
action of a secondary, intersecting the disc repeatedly while
orbiting the centre
\cite{SubK99},
oscillations of the (thin) disc itself, perturbed off its
circular (equatorial) flow
\cite{TitLM98,Wag99,Kat01},
or waves in tilted discs resulting from the Bardeen-Peterson
effect \cite{FraMW01}.
(Clearly, the processes can proceed in a symbiosis, for instance
the secondary would probably actuate hot spots as well as
oscillations in the disc, pulling some of its material out into a
tilted orbit.)
See \cite{Kli00} for a survey and
\cite{SteVM99,CuiSHH00,Str01a,Str01b,NanMRCh01} for discussions
of the individual models in the light of new observations.

Among the sources displaying the presence of an interacting
ultracompact body there is a class whose luminosity is still very
low and fluctuating within a wide range. This seems to point
towards the existence of a horizon, because a quasi-stationary
inflow onto a neutron star should yield a relatively strong and
steady output, produced mainly by the gas streams colliding with
the star's surface. With a horizon, the accretion may (perhaps
occasionally) switch over to an ``advection-dominated'' regime
when the material plunges in without having released any
significant fraction of its binding energy
\cite{NarGM97,Garetal01}.
See \cite{AbrChGL96} for a general model of advection-dominated
discs, \cite{IguA00,IguAN00,KurJ00,Man00,MachMM01} for recent
numerical simulations of their structure and spectrum,
\cite{Mei01,BecSK01} for a study of jet production in such discs,
and \cite{MinM99} for a review.

The main goal of precise measurements of sources suspected of
core collapse is a determination of three independent parameters
--- mass of the compact centre $M$, its specific angular momentum
$a$, and radius $r$ where a given process happens. Describing the
core field by the Kerr metric, the above three parameters appear
in the formulae employed in models of the observed phenomena.
These can be inverted if the necessary independent observables
(at least three, in general) are really measured; see e.g.
\cite{SemKF99} for a simple example of such an inference,
\cite{Gebetal00} for main techniques of fixing the central $M$ in
galaxies (\cite{MacB01} proposed a new one), and e.g.
\cite{BaoHWX97,MerF01,AdaGR01,LuYu01} for other prospects.

The dividing line between neutron stars and black holes being
given by the Landau-Oppenheimer-Volkoff mass limit of about
$M=(2\div 2.5)M_{\odot}$ ($M_{\odot}$ is the solar mass),
attention has recently been focused on determination of the
rotation rate of the more massive cases. There is some evidence
that highly active sources (in ``hard X-state'', ``radio-loud'',
with strong and fast jets) host rapidly rotating holes and vice
versa
\cite{ZhaCCh97,CuiZCh98,Wag99,Maketal00,Mei01,Str01a,GieME01}.
In particular, the popular microquasars GRS 1915+105 and GRO
J1655-40 favour rapid central rotation, perhaps even close to the
extreme value $a=M$ which separates black holes from objects
without a horizon (naked singularities).\footnote
{Note that GRS 1915+105 provided the first direct evidence of the
 disc-jet interaction \cite{EikMMRN98}; GRO J1655-40 is likely to
 have originated in a supernova explosion \cite{IsrRBCM99}.
 Their central spins however remain uncertain --- cf. the
 disjoint ranges estimated in \cite{GieME01}, \cite{AbrK01}
 and \cite{WagSO01}. The spin of the central black hole in our
 Galaxy is discussed in \cite{MelBLW01}.}
Such a rapid rotation is expected due to theoretical reasons ---
calculations of the effect of disc accretion on the central body
\cite{Wan00,AgoK00}. On the other hand, a hyper-extreme object
would probably be slowed down to $a=M$ by accretion
\cite{CalN79,Stu81}.\footnote
{Conversely, cf. a recently proposed gedanken experiment
 \cite{FelY01} which appears to show that it is possible, in
 principle, to turn an extreme (Reissner-Nordstr\"om) black hole
 into a (Kerr-Newman) naked singularity by a radial infall of a
 (strongly bound, extended and electrically neutral) spinning
 body. In a related example, \cite{Hub99} tried to overcharge a
 near-extremal Reissner-Nordstr\"om hole.}
The centre (and the disc) could also lose rotational energy to
the benefit of the material outflowing from the system. The
significance of the Blandford-Znajek mechanism of jet fuelling
\cite{BlaZ77,ReeBBP82} for the active galactic nuclei and X-ray
binaries is still under discussion
\cite{PunC90a,GhoA97,LivOP99,LeeLP01}
(while it is rather generally accepted for gamma-ray bursts
\cite{LeeWB00}). There also exist other extraction possibilities
(see \cite{WagD89} for a review), mainly the Punsly-Coroniti
variant of Penrose mechanism \cite{PunC90b} (contrary to the
Blandford-Znajek one, it does not require a magnetic field
threading the horizon) which was indeed noticed in numerical
simulations \cite{KoiMSK00}. Another efficient scenario has been
proposed recently \cite{Li00}.

Black holes are certainly favoured by the ``cosmic censorship''
hypothesis \cite{Pen69} which defends predictability by excluding
the naked singularities as the outcome of a realistic
gravitational collapse. However, this ``question is still very
much open'' \cite{Pen98} (also \cite{Cla93b,Haw98}), with some
amount of evidence as well as counter-indications already given
(e.g. \cite{CharC90,AbrHST92,Rud97,Gun98,JhiM00,HarM01} and
references therein) whose actual significance is debatable (e.g.
\cite{BraMM98,Wal99}).\footnote
{Black holes are sometimes said to reside where God divided by
 zero. Regardless of whether he or she used Good coordinates or
 not, this is consistent with his/her being Almighty. Certain
 current readings of the cosmic censorship conjecture (``God
 abhors a naked singularity'' \cite{Haw98}) sound somewhat
 disturbing, on the contrary.}

The experimental task of deciding between black holes and naked
singularities thus brings us back to the fundamental problems of
the theory.
It should be emphasized here that Ji\v{r}\'{\i} Bi\v{c}\'ak has
always taught us to respect the decisive validity of
observations, notwithstanding his obvious esteem for the
penetrating ``pure reasoning'' (on the back of an envelope).
Ji\v{r}\'{\i} often shows romantic affinity to astronomy and
direct contact with nature, though majority of his results
concern mathematical aspects of the theory. I heard him
mentioning the need for a balance between the ``right and left
deviation'' in one's work. As a star on the stage of our
Institute, he also alternates a glowing guitar in ``Rock around
the clock'' and a delicate violin in Vivaldi.

\subsection{Gravity of accretion discs around black holes}
\label{gravity}

It is the subject of a standard university course to discuss the
features of the Schwarzschild, Reissner-Nordstr\"om, Kerr and
Kerr-Newman metrics (see \cite{Bic00} for a review), exactly
describing isolated stationary black holes (or naked
singularities). Black holes expected in galactic nuclei and in
X-ray binaries are not isolated, however: matter is concentrated
rather than absent there. Most of the evidence for ultracompact
bodies is in fact {\em based} on their interaction with the
surroundings. In theoretical models, the external matter is
supposed to form an accretion disc around the equatorial plane of
the centre (this is justified by the approximate reflectional
symmetry of the host systems and, at least in the discs' inner
parts, by Bardeen-Peterson effect \cite{BarP75}; for later
results, see \cite{FraMW01}). Accretion is a problem governed by
complicated magneto-hydrodynamics and radiative transfer which is
usually tackled under the assumptions of smoothness,
stationarity, axisymmetry and simple parametrization of the flow.
Gravity of the disc itself is neglected, the field is thus fully
determined by the centre, either described exactly by the Kerr
metric, approximated by the Schwarzschild one, simulated by some
pseudo-Newtonian potential, or quite reduced to a Newtonian
$(-M/r)$.

The main topic of the present paper is an inclusion of the disc
gravitational field in a fully general relativistic description.
{\em Newtonian} treatments of the problem began with Ostriker's
equilibria of uniformly rotating, polytropic, self-gravitating
slender rings \cite{Ost64}. The paper opened with references to
previous centuries and concluded with remarks on extragalactic
radio sources and stability of rings, both of which were to boom
soon. Later, polytropic self-gravitating configurations with a
realistic equation of state and opacity were calculated by
\cite{KozWP79} (cold case) and \cite{FukS92,BodC92,HasEM95} (hot
case), already suggesting the possibility of a black hole present
in galactic centres. \cite{MinU96} incorporated viscosity (in
a self-similar case).\footnote
{The papers must be studied in detail in order to understand
 precisely what was done. The topic of accretion discs is very
 wide and conclusions found in the literature are often hardly
 comparable due to different kinds of simplifications made,
 relating not only to the gravitational field and symmetries, but
 mainly to the rotation law, accretion rate, equation of state,
 viscosity, heat production, radiation pressure, opacity,
 boundary conditions (material supply), etc.}
In the meantime, basic formalism for studying self-gravitating
discs around black holes in a stationary axisymmetric
{\em general relativistic} case was given by \cite{Bar73,FisM76}.

Though the mass of real inner accretion discs is usually claimed
not to exceed few percent of the central mass,\footnote
{This only applies to the ultracompact centre (cf. e.g.
 \cite{Wii82,NakOK87}, however), not to protostellar discs
 \cite{YorBL93} where even the opposite may well happen
 \cite{WooST92}.}
the role of their self-gravity is not fully understood yet. Are
any properties of accretion discs so sensitive to the details of
the field that even a small effect of the disc itself can alter
the flow significantly? It was shown by \cite{ShoW82} that the
very global structure of the disc is an example of such
properties. This paper presented an extension of the disc
``$\alpha$-models'' (both in their Newtonian version of Shakura
\& Sunyaev and in the relativistic version of Novikov \& Thorne)
which includes the effects of self-gravitation, radiation
pressure and variable opacity source. The self-consistent
configurations thus obtained were considerably thinner than their
test counterparts, showing sharp radial decrease of the vertical
dimension in the middle region. The implications of self-gravity
concerning global properties of the disc (e.g. distribution of
angular momentum and outer radius) were discussed and interpreted
in a Newtonian analysis \cite{AbrCCW84}, the roles of the disc
density and total mass were distinguished and specified in
particular. With increasing gravity of the disc, regions occurred
where free circular motion became unstable or even impossible,
thus it was stated that ``astrophysically relevant models of
thick accretion discs must incorporate self-gravity effects in
order to be self-consistent'' (the same had already been
indicated in \cite{ShoW82,Wii82} before).

Changes in the properties of circular equatorial motion were
later observed in the pseudo-Newtonian study \cite{KhaCh92} of
a rotating black hole surrounded by a massive thin equatorial
ring, in a relativistic analysis of the corresponding static case
\cite{Chak88}, and also in the perturbative solution \cite{Wil74}
for a slowly rotating, weakly gravitating thin ring around a
Schwarzschild black hole \cite{Wil75}. These changes were claimed
to result in modifications of the observable characteristics,
e.g. in higher temperature of the disc. The latter was confirmed
by numerically constructing equilibrium structures of
self-gravitating polytropic thick discs in a pseudo-Newtonian
potential \cite{BodC92}. The accretion-disc ``self-field'' was
also discussed in connection with the disc's vertical structure
and stability. It turns out that certain modes of instability can
be damped whereas other are amplified by the disc's gravity (see
\cite{PapS91,AndTCh97}). The latest fully {\em relativistic}
results --- employing numerical spacetimes obtained in
\cite{Lan92,NisE94} --- include the computation of emission-line
profiles from self-gravitating discs around rotating black holes;
the case of light, thin finite discs was treated by
\cite{KarLV95} while that of heavy toroids by \cite{UsuNE98}.

Self-gravitating accretion discs also appeared in other contexts,
some of which are interesting from the point of view of general
relativity (e.g. heavy discs around white dwarfs as a ponderable
stage of a double-white-dwarf merger \cite{HachEN86}), whereas
other are not relevant for the present review (young stars
\cite{WooST92,YorBL93}, outer parts of nuclear discs, galactic
discs \cite{ShlB87}, effect on bipolar jets \cite{FukY86}).

Recently we have studied exact solutions for a static (Bach-Weyl)
ring \cite{Chak88} and static thin annular disc \cite{LemL94}
around a non-rotating black hole; some of the results are
presented in section \ref{static}. We also checked the
possibility to generalize them to a stationary situation
\cite{ZelS00} (section \ref{stationary}) when the hole and the
disc were allowed to rotate and the self-gravity effects could be
even more pronounced. Actually, in relativity, the kinetic (in
fact {\em any}) energy also generates the field and mass-energy
currents give rise to frame dragging. Whereas mass of the hole
should be dominant in a real accretion system, the disc can bear
much (even most) of the angular momentum, thus modifying the
gravitomagnetic field of the centre significantly. This could be
important for the mechanisms of rotational energy extraction.

\subsection{Gravitational collapse and gamma-ray bursts}
\label{gamma-ray}

We should not forget the accretion system where the mass of the
disc or torus can actually be comparable to that of the central
hole. It is considered in the literature as a transient state of
gravitational collapse that could power the gamma-ray bursts.

In a present-day relativistic astrophysics, perhaps the most
popular system is a close binary of ultracompact objects. The
famous Hulse-Taylor binary pulsar (PSR 1913+16) turned out to
spiral in at the rate predicted by general relativistic loss of
angular momentum via gravitational radiation \cite{Tay94}, thus
providing the first, indirect confirmation of the existence of
gravitational waves. Nowadays late stages of evolution of such
systems (inspiral, merger and ``ringdown'') are studied
numerically or perturbatively by a number of research groups,
mainly as the most promising source of gravitational waves
\cite{Tho98}, and also as an engine suggested for the gamma-ray
bursts \cite{Mes01}. In particular, at late stages of the
neutron star--black hole binary inspiral, the neutron star
tidally breaks up. Loosely speaking, if this happens outside the
innermost stable orbit around the hole, the neutron debris
spreads into a strongly magnetized torus (otherwise it plunges
into the hole). Similar outcome is expected in case of a binary
with two neutron stars (which is expected to be less frequent
than the black hole--neutron star case \cite{BetB98}). It was
pointed out in numerical analyses that the black hole--neutron
torus system was liable to runaway instability (see
\cite{MasNE98} and references therein). Depending on the exact
shape of the created magnetosphere, the energy release is mainly
ensured by leptonic winds from pair-creation along the axis of
rotation or/and axially collimated Poynting outflow resulting
from the Blandford-Znajek mechanism \cite{LeeWB00,Put01}. In
\cite{Put01}, ``Long/short gamma-ray bursts are identified with
suspended/hyper- accretion onto rapidly/slowly rotating black
holes. \dots In long bursts, the torus is expected to radiate
most of the black hole luminosity in gravitational waves. This
predicts that long gamma-ray bursts are potentially the most
powerful burst-sources of gravitational waves in the Universe.''
(This thorough review also conjectures about connections between
gamma-ray bursts and active galactic nuclei and micro-quasars.)

The black hole with a disc/torus can also arise as the outcome
of a gravitational collapse of a massive rotating star. This was
already mentioned in \cite{NakOK87}, while more recently in
scenarios of ``failed supernovae'' from the cores of young stars,
either isolated \cite{FadW99} or in a binary (this case is
referred to as ``hypernovae'') \cite{Pac98,Broetal00}; for a
review, see discussions in \cite{Put01,BroBL01}.

\section{Stationary axisymmetric spacetimes}
\label{formulas}

It is natural first to try to include the gravitational effect of
the external source in the simplest case when the system is
stationary and axisymmetric. Stationary axisymmetric spacetimes
are of obvious astrophysical importance: they describe the
exterior of bodies like stars, galaxies or accretion discs in
equilibrium.
The issue has been exposed in many places (e.g.
\cite{Lev68,Car73,Bar73,Wil74,ButI76,Chan78,Emb84,Isl85,ChiGR93}
or chapters 17--19 of \cite{KraSHM80}); for more information
about the most important particular solutions, see
\cite{Bon92,Bic00}.

In the Weyl-Lewis-Papapetrou coordinates ($t,\rho,\phi,z$) of the
cylindrical type, the metric can be written as\footnote
{The quantities will be given in geometrized units in which
 $c=G=1$; the signature of the metric tensor $g_{\mu\nu}$ is
 ($-$+++). Greek indices run from 0 to 3 and Latin indices
 ($i$, $j$, \dots) run from 1 to 3; indices from the beginning of
 the Latin alphabet ($a$, $b$, \dots) represent cyclic
 coordinates $t$ and $\phi$. Partial differentiation is denoted
 by $\partial$ or by a subscript comma, covariant derivative is
 denoted by $\nabla$ or by a subscript semicolon.}
\begin{equation}  \label{WLP}
  {\rm d}s^{2}=-e^{2\nu}{\rm d}t^{2}
               +\rho^{2}B^{2}e^{-2\nu}
                ({\rm d}\phi-\omega{\rm d}t)^{2}
               +e^{2\lambda-2\nu}({\rm d}\rho^{2}+{\rm d}z^{2}),
\end{equation}
where the unknown functions $\nu$, $B$, $\omega$ and $\lambda$
only depend on $\rho$ and $z$; $\omega$ is interpreted as the
angular velocity of inertial space rotation with respect to
observers at rest in spatial infinity. The metric coefficients
$g_{ab}$ have invariant meaning, they can be expressed in terms
of Killing fields
$k^{\mu}=\partial x^{\mu}/\partial t$ and
$m^{\mu}=\partial x^{\mu}/\partial\phi$:
\begin{equation}
  g_{tt}\,(=-e^{2\nu}+\omega^{2}g_{\phi\phi})=k_{\iota}k^{\iota},
\end{equation}
\begin{equation}
  g_{t\phi}\,(=-\omega g_{\phi\phi})=k_{\iota}m^{\iota},
\end{equation}
\begin{equation}
  g_{\phi\phi}\,(=\rho^{2}B^{2}e^{-2\nu})=m_{\iota}m^{\iota}.
\end{equation}
The simple result for the corresponding subdeterminant,
$\det(g_{ab})=-\rho^{2}B^{2}$, implies
\begin{equation}  \label{WLPinverse}
  g^{tt}=-e^{-2\nu}, \;\;\;\;\;
  g^{t\phi}=-e^{-2\nu}\omega, \;\;\;\;\;
  g^{\phi\phi}=-e^{-2\nu}\omega^{2}+\rho^{-2}B^{-2}e^{2\nu}.
\end{equation}

\subsection{Field equations}

The Einstein equations read
\begin{equation}  \label{eqB}
  \vec{\nabla}\cdot(\rho\vec{\nabla} B)=
  8\pi\rho B(T_{\rho\rho}+T_{zz}),
\end{equation}
\begin{equation}  \label{eqomega}
  \vec{\nabla}\cdot(\rho^{2}B^{3}e^{-4\nu}\vec{\nabla}\omega)=
  -16\pi Be^{2\lambda-2\nu}T^{t}_{\phi},
\end{equation}
\begin{equation}  \label{eqnu}
  \vec{\nabla}\cdot(B\vec{\nabla}\nu)=
  \frac{1}{2}\rho^{2}B^{3}e^{-4\nu}(\vec{\nabla}\omega)^{2}+
  4\pi Be^{2\lambda}(2T^{tt}+e^{-2\nu}T),
\end{equation}
where $T\equiv T^{\sigma}_{\sigma}$ and
$\vec{\nabla}$ and $\vec{\nabla}\cdot$ stand for the gradient and
divergence in a flat three-dimensional space with coordinates
($\rho$,$\phi$,$z$); thus
$\vec{\nabla}X=(X_{,\rho},0,X_{,z})$ and
$\vec{\nabla}\cdot\vec{X}=
 \rho^{-1}[(\rho X^{\rho})_{,\rho}+(\rho X^{z})_{,z}]$
in the axially symmetric case.
Once $B$, $\omega$ and $\nu$ are known, $\lambda$ can be
integrated from equations
\[B\lambda_{,z}-B_{,z}
 +\rho\left(B_{,\rho}\lambda_{,z}+B_{,z}\lambda_{,\rho}-
            B_{,\rho z}-2B\nu_{,\rho}\nu_{,z}\right)\]
\begin{equation}  \label{lambda1}
 +\frac{1}{2}\rho^{3}B^{3}e^{-4\nu}\omega_{,\rho}\omega_{,z}
 =8\pi\rho BT_{\rho z},
\end{equation}
\[2B\lambda_{,\rho}-2B_{,\rho}
 +\rho\left[2B_{,\rho}\lambda_{,\rho}-2B_{,z}\lambda_{,z}-
            B_{,\rho\rho}+B_{,zz}-2B(\nu_{,\rho}^{2}-
            \nu_{,z}^{2})\right]\]
\begin{equation}  \label{lambda2}
 +\frac{1}{2}\rho^{3}B^{3}e^{-4\nu}
  (\omega_{,\rho}^{2}-\omega_{,z}^{2})=
  8\pi\rho B(T_{\rho\rho}-T_{zz}).
\end{equation}

Although a suitable combination of solenoidal motions can also
satisfy the assumption of axial symmetry, it is natural to expect
that the axially symmetric source just rotates in the azimuthal
direction, with an angular velocity
$\Omega={\rm d}\phi/{\rm d}t$.
The corresponding four-velocity is
\begin{equation}  \label{u}
  u^{\mu}=u^{t}(k^{\mu}+\Omega m^{\mu})=u^{t}(1,0,\Omega,0),
\end{equation}
where
$(u^{t})^{-2}=-g_{tt}-2\Omega g_{t\phi}-\Omega^{2}g_{\phi\phi}
 =e^{2\nu}(1-\hat{v}^{2})$
and $\hat{v}=\rho Be^{-2\nu}(\Omega-\omega)$ is the linear speed
with respect to the local zero-angular-momentum observer.

For a source made of ideal fluid, having a total mass-energy
density $\epsilon$ and pressure $P$ as measured in a co-moving
system, the energy-momentum tensor is
$T_{\mu\nu}=(\epsilon+P)u_{\mu}u_{\nu}+Pg_{\mu\nu}$
and the equations of motion read
$(\epsilon+P)a_{\mu}=-P_{,\mu}$,
where the four-acceleration $a_{\mu}=u_{\mu;\nu}u^{\nu}$ can, for
instance, be written as
\[a_{\mu}=-\frac{1}{2}g_{\alpha\beta,\mu}u^{\alpha}u^{\beta}
         =-\frac{g_{tt,\mu}+2\Omega g_{t\phi,\mu}+
                 \Omega^{2}g_{\phi\phi,\mu}}
                {2e^{2\nu}(1-\hat{v}^{2})}\]
\begin{equation}  \label{a,1st}
\hspace{5mm}
         =\frac{2e^{2\nu}\nu_{,\mu}+
                2(\Omega-\omega)g_{\phi\phi}\omega_{,\mu}-
                (\Omega-\omega)^{2}g_{\phi\phi,\mu}}
               {2e^{2\nu}(1-\hat{v}^{2})} \, .
\end{equation}
From equations (\ref{eqB})--(\ref{eqnu}) one then obtains
\begin{equation}  \label{eqB,fluid}
  \vec{\nabla}\cdot(\rho\vec{\nabla} B)=
  16\pi\rho Be^{2\lambda-2\nu}P,
\end{equation}
\begin{equation}  \label{eqomega,fluid}
  \vec{\nabla}\cdot(\rho^{2}B^{3}e^{-4\nu}\vec{\nabla}\omega)=
  -16\pi\rho B^{2}e^{2\lambda-4\nu}(\epsilon+P)
  \frac{\hat{v}}{1-\hat{v}^{2}} \, ,
\end{equation}
\begin{equation}  \label{eqnu,fluid}
  \vec{\nabla}\cdot(B\vec{\nabla}\nu)=
  \frac{1}{2}\rho^{2}B^{3}e^{-4\nu}(\vec{\nabla}\omega)^{2}+
  4\pi Be^{2\lambda-2\nu}
  \left[(\epsilon+P)\frac{1+\hat{v}^{2}}{1-\hat{v}^{2}}+
        2P\right],
\end{equation}
and $\lambda$ is given by
\[\lambda_{,\rho}=
  \frac{B}{(\rho B)_{,\rho}^{2}+\rho^{2}B_{,z}^{2}}
  \bigg\{
    B_{,\rho}+\rho B^{-1}(B_{,\rho}^{2}+B_{,z}^{2})\]
\[\hspace{3.7cm}
   +\rho\left[(\rho B)_{,\rho}(\nu_{,\rho}^{2}-\nu_{,z}^{2})+
              2\rho B_{,z}\nu_{,z}\nu_{,\rho}\right]\]
\begin{equation}  \label{lambda,rho,fluid}
  \hspace{3.7cm}
   -\frac{1}{4}\rho^{3}B^{2}e^{-4\nu}
    \left[(\rho B)_{,\rho}(\omega_{,\rho}^{2}-\omega_{,z}^{2})+
          2\rho B_{,z}\omega_{,z}\omega_{,\rho}\right]
  \bigg\},
\end{equation}
\[\lambda_{,z}=
  \frac{B}{(\rho B)_{,\rho}^{2}+\rho^{2}B_{,z}^{2}}
  \bigg\{
    B_{,z}+
    \rho\left[\rho B_{,z}(\nu_{,z}^{2}-\nu_{,\rho}^{2})+
              2(\rho B)_{,\rho}\nu_{,\rho}\nu_{,z}\right]\]
\begin{equation}  \label{lambda,z,fluid}
  \hspace{3.7cm}
   -\frac{1}{4}\rho^{3}B^{2}e^{-4\nu}
    \left[\rho B_{,z}(\omega_{,z}^{2}-\omega_{,\rho}^{2})+
          2(\rho B)_{,\rho}\omega_{,\rho}\omega_{,z}\right]
  \bigg\}.
\end{equation}

The unknown metric functions are subject to boundary conditions
on the horizon (if there is one), on the symmetry axis and at
spatial infinity. The metric must be regular on the horizon and
on the axis; in the case of an isolated source one requires
asymptotic flatness. The conditions were discussed by
\cite{Car73,Bar73}, for example.

\subsection{Vacuum case and the Ernst formulation}
\label{vacuum}

In the following sections, we will be interested in vacuum
solutions or in those with infinitely thin sources. In both cases
$P=0$ and the equation (\ref{eqB,fluid}) has only one solution
with a satisfactory asymptotic behaviour [namely
(\ref{B,asymp})], $B=1$. Only two of the field equations then
remain, (\ref{eqomega},\ref{eqnu}), in the form
\begin{equation}  \label{eqomega,vac}
  \vec{\nabla}\cdot(\rho^{2}e^{-4\nu}\vec{\nabla}\omega)=0,
\end{equation}
\begin{equation}  \label{eqnu,vac}
  \vec{\nabla}^{2}\nu=
  \frac{1}{2}\rho^{2}e^{-4\nu}(\vec{\nabla}\omega)^{2},
\end{equation}
and the relations (\ref{lambda,rho,fluid},\ref{lambda,z,fluid})
reduce to
\begin{equation}  \label{lambda,rho,vac}
  \lambda_{,\rho}=
  \rho(\nu_{,\rho}^{2}-\nu_{,z}^{2})-
  \frac{1}{4}\rho^{3}e^{-4\nu}
  (\omega_{,\rho}^{2}-\omega_{,z}^{2}),
\end{equation}
\begin{equation}  \label{lambda,z,vac}
  \lambda_{,z}=
  2\rho\nu_{,\rho}\nu_{,z}-
  \frac{1}{2}\rho^{3}e^{-4\nu}\omega_{,\rho}\omega_{,z}.
\end{equation}

Many other formulations of the stationary axisymmetric problem
are possible, starting from a different definition of the four
(or three) unknown metric functions. In the vacuum case the
formulation by Ernst \cite{Ern68a,Ern68b} is often considered,
where the field equations are translated into the (Ernst)
equation
\begin{equation}
 -g_{tt}\vec{\nabla}^{2}{\cal E}=(\vec{\nabla}{\cal E})^{2}
\end{equation}
for a complex (Ernst) potential
${\cal E}\equiv -g_{tt}+{\rm i}\psi$
whose imaginary part $\psi$ is given by
\begin{equation}
  \rho\psi_{,z}=g_{tt}g_{t\phi,\rho}-g_{tt,\rho}g_{t\phi},
  \;\;\;\;\;
  \rho\psi_{,\rho}=g_{tt,z}g_{t\phi}-g_{tt}g_{t\phi,z};
\end{equation}
$\lambda$ is found by a line integral as above.

\subsection{Horizon}
\label{horizon}

(See e.g. \cite{Car73,Car79,FroN98} for thorough accounts.)
A black-hole horizon is a null hypersurface below which the
spacetime is dynamical. In Weyl-Lewis-Papapetrou coordinates, it
is located where $\det(g_{ab})=-\rho^{2}B^{2}$ is zero. Also
vanishing there is the {\it lapse} (or {\it redshift factor})
$\alpha\equiv e^{\nu}$ or the magnitude of the Killing bivector
$k_{[\kappa}m_{\lambda]}$; this is clear from relations
\begin{equation}
  -2k_{[\kappa}m_{\lambda]}k^{[\kappa}m^{\lambda]}
  =-\det(g_{ab})
  =\alpha^{2}g_{\phi\phi},
\end{equation}
because $g_{\phi\phi}$ (as well as $g_{\rho\rho}$) must be
regular (and positive) on the horizon. In the vacuum case $B=1$,
so the horizon lies on the axis ($\rho=0$). The interior of the
black hole thus has to be studied in different coordinates, e.g.
in spheroidal coordinates of the Boyer-Lindquist type
($t$,$r$,$\theta$,$\phi$), introduced by the transformation
\begin{equation}  \label{BLcoord}
  \rho=\sqrt{\Delta}\sin\theta, \;\;\;\;\;
  z=(r-M)y,
\end{equation}
where $\Delta=(r-M)^{2}-k^{2}$ and $y=\cos\theta$, $M$ being a
scale parameter (it usually represents the total mass) and
$(\pm)k$ determining where the horizon reaches up along the $z$
axis. Sometimes an isotropic radial coordinate $R$ is employed,
given by
\begin{equation}  \label{R}
  \Delta=R^{2}\left(1-\frac{k^{2}}{4R^{2}}\right)^{2}.
\end{equation}
On the horizon, $\Delta=0$ and both radial coordinates are
constant, $r=M+k\equiv r_{\rm H}$ (hence, $z=ky$),
$R=k/2\equiv R_{\rm H}$
(there are more black-hole horizons in general, we mean the
outermost one here).

There are three important parameters of the horizon:
its surface area (proportional to the entropy of the black
hole), the surface gravity (proportional to the horizon
temperature) and the angular velocity relative to infinity,
\begin{equation}
  A=2\pi\int\limits_{0}^{\pi}
    \sqrt{(g_{\theta\theta}g_{\phi\phi})_{\rm H}}\,
    {\rm d}\theta, \;\;\;\;
  \kappa=|\nabla e^{\nu}|_{\rm H}, \;\;\;\;
  \omega_{\rm H}=\omega(r=r_{\rm H});
\end{equation}
$\kappa$ and $\omega_{\rm H}$ are constant all over the horizon.

For spacetimes with a symmetry lower than spherical, the question
of the ``true shape'' of the black hole, not distorted by
coordinates, is non-trivial. How is the hole influenced by
rotation, electromagnetic field or an external source? The answer
is obtained by representing the horizon as a two-dimensional
surface in a three-dimensional Euclidean space. Following
\cite{Sma73}, the two-dimensional metric is first rewritten as
\begin{equation}
  {\rm d}s^{2}=
  \frac{A}{4\pi}\,[h^{-1}(y){\rm d}y^{2}+h(y){\rm d}\phi^{2}],
\end{equation}
where $h(y)=(4\pi/A)(g_{\phi\phi})_{\rm H}$.
An isometric embedding of the two-surface $(y,\phi)$ in
I$\!$E$^{3}$ with coordinates $(X,Y,Z)$ is then given by
\begin{equation}  \label{embedd}
  X=\frac{A}{4\pi}\sqrt{h}\cos\phi, \;\;\;\;\;
  Y=\frac{A}{4\pi}\sqrt{h}\sin\phi, \;\;\;\;\;
  Z=\frac{A}{4\pi}\int\limits_{0}^{y}
    \sqrt{\frac{1}{h}\left(1-\frac{1}{4}h_{,y}^{~2}\right)}
    \;{\rm d}y.
\end{equation}
If its Gaussian curvature
\begin{equation}  \label{Gauss}
  {\cal C}_{\rm H}=-\frac{8\pi^{2}}{A^{2}}\,h_{,yy}
\end{equation}
is negative anywhere, the horizon cannot be (globally) embedded
in I$\!$E$^{3}$.
It was shown in \cite{Sma73} that rotation can flatten the
horizon considerably, the Gaussian curvature of a Kerr black hole
finally turns negative at the axis for $a/M>\sqrt{3}/2$ ($a$
denotes the angular momentum per unit $M$). Some of the black
holes swirling in galactic nuclei may even be impossible to
imagine!

\subsection{Static case}
\label{Weyl}

In the static case $\omega=0$ and (\ref{WLP}) acquires the Weyl
canonical form
\begin{equation}  \label{Weylmetric}
  {\rm d}s^{2}=-e^{2\nu}{\rm d}t^{2}
               +\rho^{2}e^{-2\nu}{\rm d}\phi^{2}
               +e^{2\lambda-2\nu}({\rm d}\rho^{2}+{\rm d}z^{2}).
\end{equation}
More precisely, the metric can be written in this way provided
that $T^{\rho}_{\rho}+T^{z}_{z}=0$ (which is fulfilled by
incoherent dust, certain electromagnetic fields or certain
infinitely thin sources, but not by a fluid with non-zero
pressure), otherwise there remains $B^{2}$ in $g_{\phi\phi}$.

Equations (\ref{eqB})--(\ref{lambda2}) reduce to the Poisson
equation
\begin{equation}  \label{eqnu,Weyl}
  \vec{\nabla}^{2}\nu\equiv
  \rho^{-1}\nu_{,\rho}+\nu_{,\rho\rho}+\nu_{,zz}=
  4\pi e^{2\lambda-2\nu}(T^{\phi}_{\phi}-T^{t}_{t})
\end{equation}
for $\nu$ and to relations
\begin{equation}  \label{lambda12,Weyl}
  \lambda_{,z}-2\rho\nu_{,\rho}\nu_{,z}=8\pi\rho T_{\rho z},
  \;\;\;\;\;
  \lambda_{,\rho}-\rho(\nu_{,\rho}^{2}-\nu_{,z}^{2})=
  4\pi\rho (T_{\rho\rho}-T_{zz})
\end{equation}
for $\lambda$.

\subsection{The axis and the equatorial plane}
\label{axis,equator}

On the symmetry axis ($\rho=0$), the regularity of metric
requires $e^{\lambda}=B$, so
\begin{equation}
  {\rm d}s^{2}=-e^{2\nu}{\rm d}t^{2}+B^{2}e^{-2\nu}{\rm d}z^{2}.
\end{equation}
The formulas can also be expected to simplify in the equatorial
plane ($z=0$) if the spacetime is reflectionally symmetric with
respect to it. This is usually the case in astrophysically
motivated considerations. As a special case, the solutions with
stronger, cylindrical symmetry (where another Killing field,
$\partial x^{\mu}/\partial z$, exists) can be mentioned, where
{\em any} of the planes $z={\rm const}$ is equatorial in this
sense; however, there is no place for black holes in these
spacetimes.

\subsection{Singularities}

From astrophysical point of view, a given solution is problematic
mainly if it contains physical singularities on or above the
horizon, if it shows bad asymptotic properties or if it does not
correspond to any realistic source. The existence of a
singularity follows --- according to its type (see
\cite{EllS77,Cla93a}) --- e.g. from the divergence of scalars
constructed from the metric tensor and its derivatives or from
the divergence of physical (tetrad) components of the Riemann
tensor. (Usually, the Kretschmann invariant
$R_{\mu\nu\rho\sigma}R^{\mu\nu\rho\sigma}$ is checked first.)
However, the spacetimes are known which do contain singularities,
although all their curvature invariants vanish. It is thus
difficult to verify the {\em non-}existence of singularities; the
only proof of regularity at a given point is to find a (local)
coordinate system in which the metric is smooth enough there.

\subsection{Analytic extension and global structure}
\label{global}

Astrophysically relevant is that part of spacetime which can
communicate with spatially remote regions, i.e. the ``region of
outer communications'' outside the horizon. General relativity is
interested in the black hole interior as well --- in fact even
there an observer can perform physical observations (and perhaps
even survive). Unfortunately, in Weyl-Lewis-Papapetrou
coordinates the metric is not defined below the horizon (this
would correspond to imaginary radius $\rho$) and one has to
extend it there. Such a task may be solvable but numerically and
hardly expectable regions may open. What is the type and shape of
the singularity under the horizon? How does it change when an
additional source is ``switched on'' above the horizon? The
answer is unknown even for quite simple superpositions (on the
other hand, in a highly symmetric case it may well be possible to
solve the dynamical problem \cite{Krt02}).

Far away from the horizon, one inquires for the asymptotic
behaviour of the field. That of an isolated stationary source
falls to zero there in a specific manner --- the spacetime is
said to be asymptotically flat (see e.g. \cite{Wal84}, chapter
11, or \cite{BeiS00}, section 3). In case of the metric
(\ref{WLP}), it must hold, for $r\rightarrow\infty$,
\begin{equation}  \label{nu,asymp}
  \nu=-\frac{{\rm total~mass}}{r}+O(r^{-2}),
\end{equation}
\begin{equation}  \label{B,asymp}
  B=1+O(r^{-2}),
\end{equation}
\begin{equation}  \label{omega,asymp}
  \omega=\frac{2\cdot({\rm total~angular~momentum})}{r^{3}}
         +O(r^{-4}),
\end{equation}
\begin{equation}  \label{lambda,asymp}
  \lambda=O(r^{-2}).
\end{equation}
The courses can be expressed, in the same manner, in terms of the
isotropic radial coordinate $R$ or in terms of the radius
$\sqrt{\rho^{2}+z^{2}}$, since $R=r-M+O(r^{-1})$\footnote
{This is clear from an explicit form of (\ref{R}),
 $R=\frac{1}{2}(r-M+\sqrt{\Delta})$, or
 $r=R+M+\frac{k^{2}}{4R}$.}
and $\sqrt{\rho^{2}+z^{2}}=r-M+O(r^{-1})$ for
$r\rightarrow\infty$.

\subsection{Circular orbits and ergosphere}
\label{circular}

The simplest type of motion in stationary axisymmetric spacetimes
is that along spatially circular orbits ($\rho={\rm const}$,
$z={\rm const}$) with steady angular velocity $\Omega$: such a
motion takes place along the symmetries, namely tangent to each
individual circular orbit is a Killing field (one speaks of
``quasi-Killing'' trajectories); physically speaking, an observer
on a circular orbit sees time-independent field around and is
thus called the stationary observer. In order that the
corresponding four-velocity (\ref{u}) lie inside the light cone,
$\Omega$ must fall within the values
\begin{equation}
  \Omega_{\stackrel{\rm max}{\scriptscriptstyle\rm min}}=
  \omega\pm\frac{\alpha}{\sqrt{g_{\phi\phi}}} \, ;
\end{equation}
on the horizon (where $\alpha=0$) the permitted range narrows
down to a single value $\omega_{\rm H}$.

Calculations involving acceleration are more difficult as they
also contain derivatives of the metric. The circular-orbit
four-acceleration (\ref{a,1st}) has at most two non-zero
components ($a_{\rho}$ and $a_{z}$); $\mu$-component is zero if
\begin{equation}  \label{Omegapm}
  \Omega=\Omega_{\pm}=
  \frac{-g_{t\phi,\mu}
        \pm\sqrt{(g_{t\phi,\mu})^{2}-g_{tt,\mu}g_{\phi\phi,\mu}}}
       {g_{\phi\phi,\mu}} \, ,
\end{equation}
where the upper/lower sign corresponds to a
``co-rotating''/``counter-rotating''\footnote
{These terms may be misleading in spacetimes with large angular
 momentum.}
orbit.

In the equatorial plane, $a_{z}=0$ due to the symmetry. If
$\Omega=\Omega_{\pm}$ (here with $\mu=\rho$), $a_{\rho}=0$ as
well and the worldline is a geodesic. Among equatorial geodesics,
three particular cases are most important: the photon, the
marginally bound and the marginally stable orbits. The photon
orbits bound the regions where circular motion is time-like;
they are given by equalities
\begin{equation}  \label{photon}
  \Omega_{\pm}=
  \Omega_{\stackrel{\rm max}{\scriptscriptstyle\rm min}}
\end{equation}
[$\mu=\rho$ being the only non-trivial index value in
(\ref{Omegapm}) now].
The marginally bound orbits limit the regions where particles on
circular orbits have lower energy than is necessary for the
existence at spatial infinity; the limiting case is given by
${\gamma}(\Omega=\Omega_{\pm})=1$, where
\begin{equation}  \label{gamma,ell}
  \gamma=-u_{t}=
  -u^{t}(g_{tt}+\Omega g_{t\phi})=u^{t}e^{2\nu}+\omega\ell
  \;\;\;\; {\rm and} \;\;\;\;
  \ell=u_{\phi}=u^{t}g_{\phi\phi}(\Omega-\omega)
\end{equation}
are specific energy and specific azimuthal angular momentum at
infinity.
The marginally stable orbits bound the regions where circular
motion is stable with respect to perturbations within the orbital
plane; they satisfy $[\ell(\Omega=\Omega_{\pm})]_{,\rho}=0$. They
play a key role in the theory of accretion discs: the matter is
``swept away'' from unstable sectors; in particular, the
innermost marginally stable orbit should represent the inner disc
rim. (We will see in section \ref{static} that perturbations
{\em perpendicular} to the orbital plane can also be important.)

A peculiar feature of rotating fields are the dragging effects.
Their history starts from the famous Newton's bucket experiment
(if not earlier, see Dole\v{z}el in this volume) and continues
notably by Mach's ideas of relativity of motion and inertia
towards Einstein and Lense \& Thirring who showed, in the early
times of general relativity, that the effect is present in this
theory (see the 1913's letter of Einstein to Mach
\cite{AE93}).\footnote
{Machian inspiration keeps vivid within contemporary relativity
 --- see, for example, the last-decade references
 \cite{JanCB92,CiuW95,LynKB95,LynBK99,KinP01} or the whole
 conference \cite{BarP95}.}
The effect has an obvious analogy in electromagnetism, where
(electric) currents generate the magnetic component of the field.
This relationship was already noticed by Einstein (see his Prague
paper of 1912 \cite{AE95}) and today the gravito-electro-magnetic
approach to rotating fields prevails.

An extreme implication of dragging is the occurrence of the
ergosphere in the vicinity of ultracompact rotating objects. In
this region $\Omega_{\rm min}>0$, thus the stationary observer
cannot be static (at rest relative to infinity), albeit he still
withstands the radial attraction. The static-limit surface,
given by $\Omega_{\rm min}=0$ and hence by $g_{tt}=0$, also
represents the set of points from where signals emitted by static
observers ($\Omega=0$) reach infinity with an infinite redshift.
Ergospheres received considerable attention (and the name from
$\epsilon\rho\gamma o\nu$) after processes had been suggested
(working only there) \cite{Pen69} by means of which the
rotational energy of the centre could be extracted (without
violating the second law of black-hole dynamics that forbids the
horizon area to decrease).

With the metric (\ref{WLP}), $g_{tt}=0$ corresponds to
$\rho B\omega=\alpha^{2}$ which can only be satisfied by
$\rho\geq 0$, so the static limit really lies outside the
horizon;\footnote
{In coordinates which also describe the black hole interior,
 {\em two} horizons and {\em two} static limits are found in
 general. We mean the outer horizon and the outer static limit
 everywhere.}
in particular, $\rho=0$ implies $\alpha^{2}=0$, so the static
limit touches the horizon at the axis.

\subsection{Sources}
\label{sources}

In order to complete the relativistic solution, one must also
describe their interior (where $T_{\mu\nu}\neq 0$). Finding a
realistic interior solution is a remarkable achievement in
general, mainly if it should match smoothly the vacuum exterior
(e.g. \cite{Kyr99,Per00,Chin00} and references therein);
this is also confirmed by the history of searching for a source
of the Kerr field (see \cite{BicL93,NeuM95} and references
therein).

The problem is much simpler if the source is infinitely thin: the
solution is then vacuum-type everywhere, with the energy-momentum
tensor $T_{\mu\nu}=g_{zz}^{-1/2}S_{\mu\nu}\delta(z)$
found from the jump of the normal field across the source like in
electrodynamics; the appropriate covariant method is known as the
Israel's formalism \cite{BarI91}. It starts from a projection of
the exterior (here vacuum) metric $g_{\mu\nu}$ onto a
three-dimensional hypersurface $S$, representing the history of
the source: the projection from one ($+$) and the other ($-$)
side of $S$ yields the induced three-metrics
\begin{equation}
  ^{\pm}g_{\hat{A}\hat{B}}=
  g_{\mu\nu}\,\!^{\pm}e_{\hat{A}}^{\mu}
            \,\!^{\pm}e_{\hat{B}}^{\nu},
\end{equation}
where $\left\{\!^{\pm}e_{\hat{A}}^{\mu}\right\}_{A=0,1,2}$ are
bases tangent to $^{\pm}S$. The hyperplanes $^{\pm}S$ are
however the back and the face of the same surface, so we can
choose $^{+}e_{\hat{A}}^{\mu}=\,\!^{-}e_{\hat{A}}^{\mu}$
and only the normals $^{+}n^{\mu}$ and $^{-}n^{\mu}$ differ.
The surface density of the energy-momentum tensor
$S_{\hat{A}\hat{B}}$ reads
\begin{equation}  \label{SAB}
  S_{\hat{A}\hat{B}}=
  (8\pi)^{-1}
  \left([K_{\hat{A}\hat{B}}]-
        g_{\hat{A}\hat{B}}[K_{\hat{C}}^{\hat{C}}]\right),
\end{equation}
where $[K_{\hat{A}\hat{B}}]$ denotes the jump of the exterior
curvature across $S$, i.e.
$[K_{\hat{A}\hat{B}}]=
 \,\!^{+}K_{\hat{A}\hat{B}}-\,\!^{-}K_{\hat{A}\hat{B}}$,
with
\begin{equation}
  ^{\pm}K_{\hat{A}\hat{B}}
 =e_{\hat{A}}^{\mu}e_{\hat{B}}^{\nu}\nabla_{\mu}\,\!^{\pm}n_{\nu}
 =\,\!-^{\pm}n_{\nu}e_{\hat{A}}^{\mu}
  \nabla_{\mu}e_{\hat{B}}^{\nu}.
\end{equation}

The case of a thin source (a disc) in a stationary axisymmetric
spacetime has notably been studied by \cite{Led98,GonL00}. In a
disc lying at $z={\rm const}$ the radial pressure is zero,
$S_{\rho\rho}=0$, and for a disc in the equatorial plane ($z=0$)
of the reflectionally symmetric spacetime $S_{zz}=0$, too. Only
$S_{t\phi}$ is non-zero off the diagonal. In the
Weyl-Lewis-Papapetrou coordinates the formula (\ref{SAB}) can be
written as (\cite{Led98}, chapters 3 and 4)\footnote
{The indices $a$, $b$, \dots only run through $t$ and $\phi$ as
 anywhere.}
\begin{equation}  \label{Sab}
  S_{ab}=-\frac{\sqrt{g_{\rho\rho}}}{8\pi}
          \left(\frac{g_{ab}}{g_{\rho\rho}}\right)_{\!\!,z}.
\end{equation}
Using (\ref{WLPinverse}) and (\ref{lambda,z,vac}), this yields
\begin{equation}
  S^{t}_{t}=
 -\frac{e^{\nu-\lambda}}{8\pi}
  \left[4\nu_{,z}(1-\rho\nu_{,\rho})-
        \rho^{2}e^{-4\nu}\omega_{,z}(\omega-\rho\omega_{,\rho})
  \right],
\end{equation}
\begin{equation}
  S^{t}_{\phi}=
 -\frac{e^{\nu-\lambda}}{8\pi}\rho^{2}e^{-4\nu}\omega_{,z},
\end{equation}
\begin{equation}
  S^{\phi}_{t}=
 -\frac{e^{\nu-\lambda}}{8\pi}
  \left[4\omega\nu_{,z}-(1+\rho^{2}e^{-4\nu}\omega^{2})\omega_{,z}
  \right],
\end{equation}
\begin{equation}
  S^{\phi}_{\phi}=
  \frac{e^{\nu-\lambda}}{8\pi}\rho
  \left[4\nu_{,\rho}\nu_{,z}-
        \rho e^{-4\nu}\omega_{,z}(\omega+\rho\omega_{,\rho})
  \right].
\end{equation}

The total mass and angular momentum of the disc are fixed by the
Komar integrals that lead to
\begin{equation}  \label{Md}
  {\cal M}
 =\frac{1}{2}\int\limits_{b}^{\infty}
  g^{tc}g_{tc,z}\rho\,{\rm d}\rho
 =-\frac{1}{4}\int\limits_{b}^{\infty}
   \frac{g_{\phi\phi}^{2}}{\rho}
   \left(\frac{g_{tt}}{g_{\phi\phi}}\right)_{\!\!,z}{\rm d}\rho,
\end{equation}
\begin{equation}  \label{Jd}
  {\cal J}
 =\frac{1}{4}\int\limits_{b}^{\infty}
  g^{tc}g_{c\phi,z}\rho\,{\rm d}\rho
 =\frac{1}{4}\int\limits_{b}^{\infty}
  \frac{g_{\phi\phi}^{2}}{\rho}\omega_{,z}{\rm d}\rho,
\end{equation}
where $\rho=b>0$ is the position of the disc inner rim. All the
$z$-derivatives are understood to be calculated in the limit
$z\rightarrow 0^{+}$.

In the event that there is a black hole present in the centre of
the disc, its contributions $M_{\rm H}$ and $J_{\rm H}$ need also
to be included; for a given type of solution, the aggregate
parameters of the spacetime can be written as the sums
$M_{\rm H}+{\cal M}$ and $J_{\rm H}+{\cal J}$
(see e.g. \cite{Car79}).
The black-hole mass is related through the Smarr formula
(\cite{Car79}, equation (6.297))
\begin{equation}  \label{MH}
  M_{\rm H}=2\omega_{\rm H}J_{\rm H}+\frac{\kappa A}{4\pi}=
            2\omega_{\rm H}J_{\rm H}+k
\end{equation}
with the angular momentum
\begin{equation}  \label{JH}
  J_{\rm H}=-\frac{1}{8}\int\limits_{0}^{\pi}
             (\Delta^{2}e^{-4\nu}\omega_{,r})_{\rm H}
             \sin^{3}\theta{\rm d}\theta.
\end{equation}

The last step is a physical interpretation of the source.
First, if
\begin{equation}
  D\equiv(S^\phi_\phi-S^t_t)^2+4S^t_\phi S^\phi_t
   =\frac{e^{2\nu-2\lambda}}{16\pi^{2}}
    (4\nu_{,z}^{~2}-\rho^{2}e^{-4\nu}\omega_{,z}^{~2})
\end{equation}
is not negative, then there exists a stationary observer with the
angular velocity
\begin{equation}
  \Omega_{\rm iso}
 =(2S^t_\phi)^{-1}\left(S^\phi_\phi-S^t_t-\sqrt{D}\right)
 =\omega-\frac{2e^{4\nu}\nu_{,z}}{\rho^{2}\omega_{,z}}+
  \sqrt{\left(\frac{2e^{4\nu}\nu_{,z}}{\rho^{2}\omega_{,z}}
        \right)^{2}-\frac{e^{4\nu}}{\rho^{2}}},
\end{equation}
with respect to whom the energy-momentum tensor assumes a
diagonal (``isotropic'') form
$S^{\mu\nu}=\hat{w}u_{\rm iso}^\mu u_{\rm iso}^\nu+
            \hat{P}v_{\rm iso}^\mu v_{\rm iso}^\nu$, where
$u_{\rm iso}^\mu=u_{\rm iso}^t(1,0,\Omega_{\rm iso},0)$ and
$v_{\rm iso}^\mu=-\rho^{-1}(\ell_{\rm iso},0,\gamma_{\rm iso},0)$
are the observer's four-velocity and unit base vector in the
$\phi$-direction, $\ell_{\rm iso}$ and $\gamma_{\rm iso}$ being
the corresponding specific angular momentum and energy at infinity
(see section \ref{circular}).\footnote
{The case of $D<0$ (not involved here) represents discs with
 non-zero heat flow. Then the energy-momentum tensor must be
 written in a more general form
 $S^{\mu\nu}=\hat{w}u_{\rm iso}^\mu u_{\rm iso}^\nu+
             \hat{P}v_{\rm iso}^\mu v_{\rm iso}^\nu+
             \hat{K}(u_{\rm iso}^\mu v_{\rm iso}^\nu+
                     v_{\rm iso}^\mu u_{\rm iso}^\nu)$
 with $\hat{K}=\sqrt{-D/2}$, and
 $\Omega_{\rm iso}=(2S^t_\phi)^{-1}(S^\phi_\phi-S^t_t)
  =\omega-\frac{2e^{4\nu}\nu_{,z}}{\rho^{2}\omega_{,z}}$.
 See \cite{GonL00}.}
The observer measures the surface density
\begin{equation}
  \hat{w}=
  \frac{\gamma_{\rm iso}^{2}S^{tt}-
        \ell_{\rm iso}^{2}S^{\phi\phi}}
       {u_{\rm iso}^{t}
        (\gamma_{\rm iso}+\Omega_{\rm iso}\ell_{\rm iso})}
\end{equation}
and tangential pressure
\begin{equation}
  \hat{P}=\frac{\rho^{2}u_{\rm iso}^{t}
                (S^{\phi\phi}-\Omega_{\rm iso}^{2}S^{tt})}
               {\gamma_{\rm iso}+\Omega_{\rm iso}\ell_{\rm iso}}.
\end{equation}
If $\hat{w}\geq\hat{P}>0$, the energy-momentum tensor can
represent two equal streams of particles moving on (accelerated)
circular orbits in opposite directions at the same speed
$\sqrt{\hat{P}/\hat{w}}$. This ``speed of sound'' was shown in
\cite{GonL00} to be equal to the geometric mean of the local
prograde and retrograde circular geodesic speeds $\hat{v}_{\pm}$
as measured by the observer $u_{\rm iso}^{\mu}$:
\begin{equation}
  |\hat{v}_{+}\hat{v}_{-}|=\hat{P}/\hat{w},
\end{equation}
where
\begin{equation}
  \hat{v}_{\pm}=
  \frac{(v_{\rm iso})_t+(v_{\rm iso})_\phi\Omega_{\pm}}
       {(u_{\rm iso})_t+(u_{\rm iso})_\phi\Omega_{\pm}}
\end{equation}
and $\Omega_{\pm}$ are the angular velocities of the prograde and
retrograde circular geodesics (\ref{Omegapm}).

It was shown in \cite{Led98} that in the absence of radial
pressure the source disc can be composed of two counter-rotating
streams of non-interacting particles on time-like circular
{\em geodesics}, provided that the following conditions are
satisfied:
both geodesic frequencies (\ref{Omegapm}) lie within the
sub-luminal interval ($\Omega_{\rm min}$,$\Omega_{\rm max}$),
the tensor (\ref{Sab}) yields $S_{ab}X^{a}X^{b}\geq 0$ for at
least one two-vector $X^{a}$, and $\det(S_{ab})>0$.
The last two assumptions ensure that the superposition
\begin{equation}
  S^{ab}=w_{+}u_{+}^{a}u_{+}^{b}+w_{-}u_{-}^{a}u_{-}^{b}
\end{equation}
of the two dust components on circular geodesics
$u_{\pm}^{a}=u_{\pm}^{t}(1,\Omega_{\pm})$
comes out with positive proper surface densities of the streams
\begin{equation}
  w_{\pm}=
  \pm\frac{S^{t\phi}-\Omega_{\mp}S^{tt}}
          {(u_{\pm}^{t})^{2}(\Omega_{+}-\Omega_{-})} \, .
\end{equation}

In a static case, the above formulae simplify to
$\Omega_{\rm iso}=0$,
\begin{equation}
  \hat{w}=\frac{2w_{\pm}}{\sqrt{1-\hat{v}_{\pm}^{2}}}=
 -S^{t}_{t}=
  \frac{e^{\nu-\lambda}}{2\pi}(1-\rho\nu_{,\rho})\nu_{,z},
  \;\;\;\;\;
  \hat{P}=
  S^{\phi}_{\phi}=
  \frac{e^{\nu-\lambda}}{2\pi}\rho\nu_{,\rho}\nu_{,z},
\end{equation}
\begin{equation}
  {\cal M}=\int_{b}^{\infty}\nu_{,z}\rho{\rm d}\rho, \;\;\;\;\;
  {\cal J}=0, \;\;\;\;\;
  M_{\rm H}=k=M, \;\;\;\;\;
  J_{\rm H}=0,
\end{equation}
\begin{equation}
  \hat{v}_{\pm}=\rho e^{-2\nu}\Omega_{\pm}=
  \pm\sqrt{\frac{\rho\nu_{,\rho}}{1-\rho\nu_{,\rho}}}=
  \pm\sqrt{\hat{P}/\hat{w}} \, .
\end{equation}
Hence, the disc particles circumscribe circular geodesics in the
spacetime which they generate themselves.

Now further physical requirements can be raised, namely the weak
energy condition
($\hat{w}\geq 0$, $\hat{P}\geq -\hat{w}$),
the dominant energy condition
($\hat{w}\geq 0$, $|\hat{P}|\leq\hat{w}$),
and the non-negativity of pressure ($\hat{P}\geq 0$).
Their combination yields $\hat{w}\geq\hat{P}\geq 0$.
The requirements are very simple in the static case:
if $\nu_{,z}(z=0^{+})>0$ (which is fulfilled for a realistic
matter), then $\hat{P}\geq 0$ is equivalent to
$\nu_{,\rho}\geq 0$ which is satisfied below the Lagrangian point
of zero field (where $\nu_{,\rho}=0$ and $\hat{v}_{\pm}^{2}=0$).
In regions with tension ($\hat{P}<0$), the particles are more
attracted by the outer parts of the disc than by the central
black hole, thus no circular geodesics exist (then hoop stresses
have to be employed to interpret the disc matter); this typically
happens in discs with too much matter on larger radii.
The other condition $\hat{w}\geq\hat{P}$ is equivalent to
$\hat{v}_{\pm}^{2}\leq 1$.
Hence, the above constraints are in fact ensured by the obvious
condition $0\leq\hat{v}_{\pm}^{2}\leq 1$ which can be written
explicitly as $1\geq 1-\rho\nu_{,\rho}\geq 1/2$.

\section{Static thin discs around non-rotating black holes}
\label{static}

In a static spacetime rotation is excluded or it has to be
compensated exactly as in the case of counter-rotating streams of
matter. Within the vacuum outside of the sources, the static
axisymmetric Einstein equations (\ref{eqnu,Weyl}) and
(\ref{lambda12,Weyl}) yield the Laplace equation for $\nu$ and a
simple quadrature for $\lambda$ (calculated along a vacuum path
going from the axis to the given point),
\begin{equation}
  \vec{\nabla}^{2}\nu=0, \;\;\;\;\;
  \lambda=
  \int\limits_{\rm axis}^{\rho,z}
  \rho\,[(\nu_{,\rho}^{2}-\nu_{,z}^{2})\,{\rm d}\rho
         +2\nu_{,\rho}\nu_{,z}{\rm d}z].
\end{equation}
Thanks to the linearity of the Laplace equation, the solutions
can simply be added to obtain fields of multiple sources.
However, such a superposition may not be physically acceptable
--- supporting singularities (``struts'') may occur in
calculating $\lambda$ or it may not be possible to interpret the
resulting system of sources in a reasonable way (for instance,
the matter works out with negative density or pressure or moving
at a superluminal speed). Consequently, even the static
axisymmetric case has only afforded a few realistic
superpositions. Let us refer to two of them involving a
Schwarzschild black hole: the superposition with an infinite
annular thin disc, obtained by Lemos \& Letelier \cite{LemL94} by
inversion of the first counter-rotating finite disc of Morgan \&
Morgan \cite{MorM69}, and the one with an inverted isochrone thin
disc, obtained by Klein \cite{Kle97}. The Lemos-Letelier solution
has turned out to allow for physically satisfactory situations.
We briefly summarize its properties below.

\subsection{Superposition with the inverted first Morgan-Morgan
            disc}
\label{sequence}

The inverted first Morgan-Morgan counter-rotating disc has a
surface density
\begin{equation}
  w(\rho)=\frac{2{\cal M}b}{\pi^{2}\rho^{4}}
          \sqrt{\rho^{2}-b^{2}} \, ,
\end{equation}
${\cal M}$ and $b$ denoting the mass and the Weyl inner radius of
the disc. The matter is thus concentrated near the rim ($w$ is
maximal at $\rho=2b/\sqrt{3}\doteq 1.115b$) which is supposed to
be the case in real accretion discs. The resulting spacetime is
given by a sum of potentials which in Weyl coordinates read
\begin{equation}
  \nu_{\rm Schw}=
  \frac{1}{2}\ln\frac{d_{1}+d_{2}-2M}{d_{1}+d_{2}+2M}
\end{equation}
and
\[\nu_{\rm disc}=
 -\frac{{\cal M}}{\pi(\rho^{2}+z^{2})^{3/2}}
  \left[\left(2\rho^{2}+2z^{2}-b^{2}\frac{\rho^{2}-2z^{2}}
                                         {\rho^{2}+z^{2}}\right)
        {\rm arccot}\sqrt{\frac{\sigma-(\rho^{2}-b^{2}+z^{2})}
                               {2(\rho^{2}+z^{2})}}\right.\]
\begin{equation}  \label{nudisc}
\left. \hspace{4.2cm}
 -(3\sigma-3b^{2}+\rho^{2}+z^{2})
  \sqrt{\frac{\sigma-(\rho^{2}-b^{2}+z^{2})}
             {8(\rho^{2}+z^{2})}}
  \right],
\end{equation}
where
$d_{1,2}=\sqrt{\rho^{2}+(z\mp M)^{2}}$ and
$\sigma=\sqrt{(\rho^{2}-b^{2}+z^{2})^{2}+4b^{2}z^{2}}$;
$M$ is the Schwarzschild mass.
On the axis ($\rho=0$), $\nu_{\rm disc}$ reduces to
\begin{equation}
  \nu_{\rm disc}=
 -\frac{2{\cal M}}{\pi|z|^{3}}
  \left[(z^{2}+b^{2})\arctan\frac{|z|}{b}-b|z|\right],
\end{equation}
while in the equatorial plane ($z=0$)
\begin{equation}  \label{nu,above}
  \nu_{\rm disc}(\rho>b)=
 -\frac{{\cal M}}{\rho}
  \left(1-\frac{b^{2}}{2\rho^{2}}\right),
\end{equation}
\begin{equation}  \label{nu,below}
  \nu_{\rm disc}(\rho<b)=
 -\frac{{\cal M}}{\pi\rho}
  \left[\left(2-\frac{b^{2}}{\rho^{2}}\right)
        \arcsin\frac{\rho}{b}+
        \sqrt{\frac{b^{2}}{\rho^{2}}-1}\,\right].
\end{equation}

In \cite{SemZZ99a} we illustrated the shape of the superposed
field by drawing its field-lines, namely the integral curves of
the four-acceleration $a^{\mu}=\nabla^{\mu}\nu$ of a static
congruence. Then we plotted the flattening of the black hole with
increasing relative mass of the disc or with decreasing inner
disc radius. The explicit (static axisymmetric) form of the
horizon Gaussian curvature (\ref{Gauss}),
\begin{equation}
  {\cal C}_{\rm H}=
  \frac{1-4y\nu_{{\rm ext},y}-
        (1-y^{2})(2\nu_{{\rm ext},y}^{~2}-\nu_{{\rm ext},yy})}
       {4M^{2}e^{2\nu_{\rm ext}(y)-4\nu_{\rm ext}(1)}}
\end{equation}
(the exterior potential $\nu_{\rm ext}$ is represented by
$\nu_{\rm disc}$ in our case),
implies that starting from certain (though unrealistic) values,
${\cal C}_{\rm H}$ becomes negative at the axis.

The next paper \cite{SemZZ99b} dealt with the motion of free test
particles in the superposed fields, especially with the influence
of the disc parameters on the positions of important equatorial
circular geodesics (the photon, the marginally bound and the
marginally stable one) which are critical for the disc accretion
flows. The following questions arose: with increasing disc mass,
how does the inner rim have to shift if the whole disc is to be
stable permanently? Does it shift towards or away from the
horizon starting from the Schwarzschild radius $r=6M$ where
the marginally stable orbit lies in the case of a {\em test} disc
(${\cal M}=0$)? How do other properties of the resulting
spacetime evolve? The answers were given in papers
\cite{SemZ00a,SemZ00b,ZacS01}.

A sequence of superposed solutions was generated (figure 4 in
\cite{SemZ00a}) by gradually increasing the relative disc mass
${\cal M}/M$, while demanding that (i) all the disc matter can be
interpreted as two equal counter-rotating streams of particles on
stable time-like equatorial circular geodesics and that (ii) the
inner disc rim is fixed at the smallest possible radius. Keeping
the rim at the marginally stable orbit of the complete spacetime
turned out not to be the correct recipe for maintaining the
{\em whole} disc stable while changing the parameters; namely an
instability first appears {\em inside} the disc, not at the rim.
With increasing disc mass, the rim may first shift towards the
horizon:\footnote
{Such a statement is of course coordinate-dependent. Therefore,
 we also used some physically more relevant measures rather then
 the Schwarzschild radius $r$ only --- the equatorial
 circumferential radius
 $R=\rho e^{-\nu(\rho,z=0)}
   =re^{-\nu_{\rm disc}(r,\theta=\pi/2)}$
 and the proper radial distance from the horizon (computed along
 the equatorial plane)
 $d_{\rho}
  =\int_{0}^{\rho}e^{(\lambda-\nu)(\rho,z=0)}{\rm d}\rho
  =\int_{2M}^{r}
   \frac{e^{(\lambda-\lambda_{\rm Schw}-\nu_{\rm disc})
            (r,\theta=\pi/2)}}
        {\sqrt{1-2M/r}}{\rm d}r$.
 Note that these transformations are non-trivial as they involve
 the functions $\nu_{\rm disc}$ and $\lambda_{\rm disc}$.
 Consequently, a given $r$ corresponds, for different discs,
 to {\em different} $R$'s and $d_{\rho}$'s (and vice versa).
 However, the graphs using $r$ and using physical radii turned
 out not to differ much (see the direct comparison of all the
 three coordinates in figure 7 of \cite{SemZ00a}).}
the minimal possible Schwarzschild radius $r$ of the rim goes
from $6.00M$ to $3.60M$ with ${\cal M}$ growing from 0 to
$0.42M$;
the minimal possible circumferential radius $R$ of the rim goes
from $6.00M$ to $3.86M$ with ${\cal M}$ growing from 0 to
$0.29M$;
and the minimal possible proper distance from the horizon
$d_{\rho}$ of the rim goes from $7.19M$ to $4.26M$ with
${\cal M}$ growing from 0 to $0.33M$.
Along the generated sequence of stable discs with minimal inner
radii, the radius of the marginally bound orbit below the disc
was also found to decrease, whereas the photon orbit went slowly
up; the horizon inflated gradually towards the external source.
For ${\cal M}$ greater than the values given above, the inner
radii of the sequence increase again. When the disc mass reaches
$1.92M$, the position of the inner rim starts to be constrained
by the Lagrangian point of zero field (below which free circular
motion is impossible within a certain radial range); this,
however, goes up steeply to infinity at ${\cal M}=2M$, i.e. the
solutions of the given type with ${\cal M}\geq 2M$ cannot be
given the freely counter-rotating interpretation. The Keplerian
orbital speed was checked to be subluminal everywhere within the
resulting sequence of discs, reaching roughly 0.7 at its maximum.

Only the beginning of the above sequence of stable physical discs
can be realistic astrophysically because the mass ${\cal M}$ of
the actual accretion discs is not supposed to exceed
$(0.01\div 0.1)M$.
For ${\cal M}$ between 0 and $0.0722M$, the inner rim of our disc
sequence follows the marginally stable circular geodesic of the
corresponding complete spacetime (for ${\cal M}>0.0722M$, the rim
has to lie {\em above} the marginally stable orbit to avoid
instability of the disc at larger radii). Even with such a minor
increase of mass, the rim can go down considerably:
from $r=6M$ to $r=4.6162M$, from $R=6M$ to $R=4.6644M$ and
from $d_{\rho}=7.1914M$ to $d_{\rho}=5.4806M$.

\subsection{Oscillations and stability of self-gravitating discs}
\label{oscill}

The basic criterion for accretion-disc stability,
$\ell_{,\rho}>0$, only concerns perturbations within the disc
plane. Actually, the frequency $\kappa$ (measured with respect to
infinity) of a small free horizontal oscillation about a
circular equatorial geodesic in a stationary axisymmetric
spacetime is given by\footnote
{Note that we consider non-gravitating perturbations, i.e. the
 metric itself is not perturbed. It would be more sophisticated,
 and also much more difficult, to allow the perturbation to
 generate its own, self-consistent field. However, this would
 only have a second-order effect on oscillation frequencies.}
\begin{equation}
  \kappa^{2}
 =\frac{e^{2\nu-2\lambda}}{(u^{t})^{3}\ell}\,
  g_{\alpha\phi,\rho}u^{\alpha}g^{t\beta}u_{\beta,\rho}
 =\frac{e^{2\nu-2\lambda}\ell_{,\rho}}{(u^{t})^{4}\rho^{2}B^{2}}
  \left[\ell_{,\rho}-(u^{t})^{3}\rho^{2}B^{2}\Omega_{,\rho}
       \right]
\end{equation}
(\cite{SemZ00b} and references therein).
These oscillations (called epicyclic) have been studied in the
astrophysical literature since the 1980's, recently as the cause
of low-frequency disc modes that might explain the quasi-periodic
variability observed at some sources probably containing a black
hole with an accretion disc (see section \ref{AGN} and the review
\cite{Kat01}; the history of frequency formulae is described in
\cite{SemZ00b}). $\kappa$ is zero at radial infinity; in the
Newtonian case it increases monotonically towards the centre,
whereas in relativity it has a maximum at a certain distance and
then goes back to zero at the marginally stable orbit. Hence, in
black-hole sources, the disc oscillations produced near the inner
rim should remain trapped in a region below this maximum rather
than propagate to the outer parts of the disc \cite{KatF80}.

In \cite{SemZ00b}, perturbations in vertical direction were found
to be also important, in fact, to be even more dangerous than the
horizontal ones for discs with a greater mass. Perturbations
perpendicular to the disc plane must be treated carefully because
they take place in the vicinity of the source which is singular
(infinitely thin). Although the symmetry would imply
$g_{\mu\nu,z}=0$ in the equatorial plane, there appears a jump of
the field in the normal direction as is usual in case of a mass
(charge) layer. For the inverted first Morgan-Morgan disc, in
particular, one finds that (at $\rho>b$)
\begin{equation}  \label{limit}
  \lim_{z\rightarrow 0^{\pm}}\nu_{{\rm disc},z}=
  \pm \frac{4{\cal M}b}{\pi\rho^{4}}\sqrt{\rho^{2}-b^{2}} \, .
\end{equation}
Even in ``small'' (linear) oscillations the particle spends all
of the time {\em outside} of the equatorial plane where the field
is given by (\ref{limit}) rather than vanishing. The equation of
geodesic deviation then yields
\[\Omega_{\perp}^{2}=
  \Gamma^{z}_{~tt,z}+2\Gamma^{z}_{~t\phi,z}\Omega+
  \Gamma^{z}_{~\phi\phi,z}\Omega^{2}\]
\begin{equation}
\hspace{1cm}
 -4(\Gamma^{z}_{~tt}+\Gamma^{z}_{~t\phi}\Omega)
   (\Gamma^{t}_{~tz}+\Gamma^{t}_{~\phi z}\Omega)
 -4(\Gamma^{z}_{~t\phi}+\Gamma^{z}_{~\phi\phi}\Omega)
   (\Gamma^{\phi}_{~tz}+\Gamma^{\phi}_{~\phi z}\Omega)
\end{equation}
for the ``perpendicular'' frequency.

In the static case, the frequency formulae simplify to
\begin{equation}  \label{kappa,Weyl}
  \kappa^{2}=\frac{e^{4\nu-2\lambda}}{1-\rho\nu_{,\rho}}\,
             (\nu_{,\rho\rho}+4\rho\nu_{,\rho}^{3}-
              6\nu_{,\rho}^{2}+3\nu_{,\rho}/\rho),
\end{equation}
\begin{equation}  \label{Omegaperp}
  \Omega_{\perp}^{2}=
  \frac{e^{4\nu-2\lambda}}{1-\rho\nu_{,\rho}}
  \left[\nu_{,zz}-4\nu_{,z}^{2}(1-2\rho\nu_{,\rho})\right].
\end{equation}
Without the external source, $\Omega_{\perp}^{2}$ reduces to the
square of the Schwarzschild orbital frequency,
$\Omega_{\pm}^{2}=M(M+\sqrt{\rho^{2}+M^{2}})^{-3}$,
whereas $\kappa^{2}$ remains different due to the pericentre
precession effect.

Requiring linear stability (also) with respect to vertical
perturbations, the range of physically acceptable Lemos-Letelier
superpositions within the (${\cal M},b$)-plane is narrowed
further (figure 1 in \cite{ZacS01}). The curve of the marginal
vertical stability is more restrictive than that of the marginal
horizontal stability for the discs with ${\cal M}>0.2296M$, it
even fully excludes the discs with ${\cal M}\geq 2M/7$. As the
boundary of the delimited sector of physical discs (figure 2 in
\cite{ZacS01}), one obtains a sequence of stable discs whose
inner rims lie right on, or very close to, the circular geodesics
marginally stable with respect to both perturbations. We learned,
in particular, that for the ``realistic'' part of this sequence,
with relative mass below 0.07, the self-gravity makes the
oscillations of the inner disc parts faster, and that the
region of horizontal mode trapping gets somewhat smaller.

\subsection{Redshift from observers in the disc}
\label{redshift}

The redshift from a given point in the source is often mentioned
as a measure of the field strength. In Weyl fields, the frequency
shift between an observer moving on a circular orbit with an
angular velocity $\Omega$ and an observer at rest at infinity
reads
\begin{equation}  \label{g}
  g\equiv\frac{f_{\infty}}{f_{\rm emit}}=
  e^{\nu}\sqrt{1-e^{-4\nu}\rho^{2}\Omega^{2}}.
\end{equation}
In our case, it is natural to compute the shift from two
privileged observers within and below the disc, the static one
($\Omega=0$) and the geodesic one
($|\Omega|=\frac{e^{2\nu}}{\rho}
           \sqrt{\frac{\rho\nu_{,\rho}}{1-\rho\nu_{,\rho}}}$;
the observer co-rotates with the matter in terms of which the
disc is interpreted). Figure 3 of \cite{ZacS01} shows the
curves obtained (for discs with different mass and inner radius)
from the respective forms of equations (\ref{g}), $g=e^{\nu}$ and
$g=e^{\nu}\sqrt{\frac{1-2\rho\nu_{,\rho}}{1-\rho\nu_{,\rho}}}$,
by varying the emitter's orbital radius.

\subsection{Singularity at the rim of the first Morgan-Morgan
            disc}
\label{singul}

The class of Morgan-Morgan \cite{MorM69} static axisymmetric
solutions describes the fields of a sequence of finite thin
counter-rotating discs with densities
\begin{equation}  \label{density}
  w^{(m)}(\rho\leq b)=
  \frac{(2m+1){\cal M}}{2\pi b^2}
  \left(1-\frac{\rho^{2}}{b^{2}}\right)^{\!m-1/2} \;\;\;\;\;
  (m=1,2,\dots).
\end{equation}
The zeroth ($m=0$) member being clearly singular at the rim, the
main attention has been devoted to the first ($m=1$) member.
The latter has been discussed or referred to as a prototype of a
simple and physically meaningful Weyl field. In addition to the
papers mentioned in previous sections, the first Morgan-Morgan
disc has appeared as the limiting case of a more general class of
{\em stationary} disc solutions \cite{Ans01,FraK01}, while its
inverted counterpart has been considered as a possible seed for
stationary black hole--disc superpositions \cite{ZelS00} (see the
next section).

The first Morgan-Morgan disc has however turned out to have a
curvature singularity at the rim \cite{Sem01}. It is caused by
too steep a decrease of density (\ref{density}): the gradient
\begin{equation}
  w^{(m)}_{,\rho}=
 -\frac{(4m^{2}-1){\cal M}\rho}{2\pi b^4}
  \left(1-\frac{\rho^{2}}{b^{2}}\right)^{\!m-3/2}
\end{equation}
goes to infinity at $\rho\rightarrow b^{-}$ if $m=0,1$.
The singularity is inherited by the annular discs, obtained by
the inversion
\begin{equation}  \label{inversion}
  \rho\rightarrow R=\frac{b^{2}\rho}{\rho^{2}+z^{2}}, \;\;\;\;\;
  z\rightarrow Z=\frac{b^{2}z}{\rho^{2}+z^{2}},
\end{equation}
(see \cite{ZacS01}).
It was thus recommended \cite{Sem01} to consider ``higher''
($m>1$) members of the Morgan-Morgan counter-rotating family in
superpositions. More accurately, even these discs have certain
singularities at their rims --- those calculated from higher
derivatives of the metric. This can already be expected from a
``Newtonian look'' at densities (\ref{density}): the $m$-th
member of the family has infinite $m$-th derivative of density at
the rim.

\subsection{Inverting the $m$-th Morgan-Morgan counter-rotating
            disc}
\label{higher}

The potential for the whole family can be most easily written in
oblate spheroidal coordinates $x$ and $y(=\cos\theta)$,
introduced by
\begin{equation}  \label{oblate}
  \rho^{2}=b^{2}(x^{2}+1)(1-y^{2}), \;\;\;\;\;
  z=bxy \;\;\;\;\;
  (0\leq x<\infty,\;1\geq y\geq -1):
\end{equation}
\begin{equation}  \label{nu(m)}
  \nu_{\rm MM}^{(m)}=
 -\frac{{\cal M}}{b}
  \sum_{n=0}^{m}C^{(m)}_{2n}{\rm i}Q_{2n}({\rm i}x)P_{2n}(y),
\end{equation}
where
\begin{equation}  \label{C2n}
  C^{(m)}_{2n}=(-1)^{n}\frac{(4n+1)(2n)!(2m+1)!(m+n)!}
                            {(n!)^{2}(m-n)!(2m+2n+1)!} \;\;\;\;\;
  (n\leq m),
\end{equation}
$P_{2n}(y)$ and $Q_{2n}({\rm i}x)$ being the Legendre polynomials
and Legendre functions of the second kind, respectively;
note that it is often suitable to express
\begin{equation}
  {\rm i}Q_{2n}({\rm i}x)=
  P_{2n}({\rm i}x){\rm arccot}\,x-
  {\rm i}\sum_{k=1}^{2n}\frac{1}{k}P_{k-1}({\rm i}x)
                                   P_{2n-k}({\rm i}x).
\end{equation}

Thanks to the invariance of the Laplace equation with respect to
the Kelvin transformation, one can invert finite thin discs with
respect to their outer rims, to obtain annular discs (which can
in principle be superposed with some central body then).
In terms of the oblate coordinates, the inversion
(\ref{inversion}) is
\begin{equation}
  x\rightarrow X=\frac{y}{\sqrt{x^{2}+1-y^{2}}}, \;\;\;\;\;
  y\rightarrow Y=\frac{x}{\sqrt{x^{2}+1-y^{2}}},
\end{equation}
where $X$, $Y$ are assumed to be related to $R$, $Z$ in the same
manner (\ref{oblate}) as $x$, $y$ are related to $\rho$, $z$.
The inverted counterpart of the solution
$\nu_{\rm MM}^{(m)}(x,y)$,
satisfying the Laplace equation and having the correct
asymptotics $\sim-\frac{{\cal M}}{bx}$, reads
\begin{equation}  \label{nu(m)inv}
  \nu^{(m)}_{\rm disc}=
  \frac{2^{2m+1}(m!)^{2}}{\pi(2m+1)!}
  \frac{\nu_{\rm MM}^{(m)}(X,Y)}{\sqrt{x^{2}+1-y^{2}}}.
\end{equation}
The inverted discs of the Morgan-Morgan class can now be
superposed (for example) with a Schwarzschild black hole or they
can be chosen as seeds in some procedure generating
{\em stationary} generalization of such a superposition (as in
the two-soliton inverse-scattering solution described in the
following section).

\section{Stationary thin discs around rotating black holes}
\label{stationary}

In order to proceed to stationary superpositions, we employed
the Belinskii-Zakharov inverse-scattering method
\cite{BelZ78,BelZ79} --- one of the generating techniques which,
in principle, provide very general classes of solutions for the
problem with two commuting symmetries. In \cite{ZelS00}, a
real-two-soliton version of the method was applied to a general
Weyl metric (\ref{Weylmetric}); the metric functions of this
``seed'' will be denoted by a hat, i.e. $\hat{\nu}$ and
$\hat{\lambda}$, leaving $\nu$ and $\lambda$ (and $\omega$) for
the {\em resulting} potentials. In Boyer-Lindquist--type
coordinates ($t$,$r$,$\theta$,$\phi$), introduced by
(\ref{BLcoord}), the obtained metric appears as
\[{\rm d}s^{2}=
  -\frac{\Delta}{\Sigma}
   \left({\cal P}e^{\hat{\nu}}{\rm d}t+
         {\cal S}e^{-\hat{\nu}}{\rm d}\phi\right)^{2}
  +\frac{\sin^{2}\theta}{\Sigma}
   \left({\cal R}e^{\hat{\nu}}{\rm d}t-
         {\cal T}e^{-\hat{\nu}}{\rm d}\phi\right)^{2}\]
\begin{equation}  \label{metric}
\hspace{1.1cm}
  +\bar{C}^{2}e^{2\hat{\lambda}-2\hat{\nu}}
   \left(\frac{\Sigma}{\Delta}{\rm d}r^{2}+
         \Sigma{\rm d}\theta^{2}\right).
\end{equation}
It involves two functions given by quadratures and it depends on
(two) potentials describing the seed spacetime and on five
independent constants; one can also choose two sign values in the
derivation. If the seed is flat, one ends up with the Kerr-NUT
solution. Restricting to reflectionally symmetric, asymptotically
flat spacetimes containing a black hole and adjusting the
coordinates in the most natural manner, the only freedom remains
in the seed ($\hat{\nu}$, $\hat{\lambda}$) and in two constants
(one puts $\bar{C}^{2}=1$, among others) which we denote $M$ and
$a$ as in the Kerr solution. The functions present in
(\ref{metric}) then read
\begin{equation}
  \Delta=(r-M)^{2}-k^2=r^{2}-2Mr+a^{2},
\end{equation}
\begin{equation}  \label{Sigma}
  \Sigma=\left(r{\cal P}+\frac{a^{2}}{k}\sinh u\right)^{2}+
         ({\cal R}y+a\sinh v)^{2}
  \;\;\;\; (={\cal P}{\cal T}+{\cal R}{\cal S}),
\end{equation}
\begin{equation}
  {\cal P}=\cosh u-(M/k)\sinh u,
\end{equation}
\begin{equation}
  {\cal R}=a\cosh v,
\end{equation}
\begin{equation}  \label{S}
  {\cal S}=(1+y^{2}){\cal R}+2ay\sinh v-2a{\cal P}e^{2\hat{\nu}},
\end{equation}
\begin{equation}
  {\cal T}=(r^{2}-a^{2}){\cal P}+\frac{2a^{2}}{k}(r-M)\sinh u+
           2a{\cal R}e^{2\hat{\nu}},
\end{equation}
where $k\equiv\sqrt{M^{2}-a^{2}}$ and $u$ and $v$ are given by
equations
\begin{equation}  \label{uv1}
  v_{,\theta}=-\frac{\Delta u_{,r}}{k\sin\theta}
             =\frac{2\Delta}{\Delta+k^{2}\sin^{2}\theta}
              \left[(r-M)\hat{\nu}_{,r}\sin\theta+
                    \hat{\nu}_{,\theta}y\right],
\end{equation}
\begin{equation}  \label{uv2}
  v_{,r}=\frac{u_{,\theta}}{k\sin\theta}
        =-\frac{2}{\Delta+k^{2}\sin^{2}\theta}
          \left[(r-M)\hat{\nu}_{,\theta}\sin\theta-
                \Delta\hat{\nu}_{,r}y\right],
\end{equation}
with boundary conditions
$u(y=\pm1)=0$, $v(y=\pm1)=\pm2\hat{\nu}(|y|=1)$.

A number of properties of the above metric were analyzed. The
(outer) horizon is given by (the larger root of) the equation
$\Delta=0$. From equations of section \ref{horizon}, its area
works out at
\begin{equation}  \label{A}
  A=4\pi\left(r_{\rm H}^{2}e^{-2\hat{\nu}_{\rm H}(y=1)}+
              a^{2}e^{2\hat{\nu}_{\rm H}(y=1)}\right)
\end{equation}
and its surface gravitation and angular velocity with respect to
infinity at
\begin{equation}  \label{kappa,omegaH}
  \kappa=\frac{4\pi k}{A}
  \;\;\;\;\; {\rm and} \;\;\;\;\;
  \omega_{\rm H}=
  \frac{4\pi a}{A}\cosh 2\hat{\nu}_{\rm H}(y=1).
\end{equation}
The Gaussian curvature of the horizon reduces to
\begin{equation}  \label{CH}
  {\cal C}_{\rm H}(y=1)=
  \frac{\left(r_{\rm H}^{2}-a^{2}e^{4\hat{\nu}_{\rm H}(1)}
             \right)
        [1-4\hat{\nu}_{{\rm H},y}(1)]
        -2a^{2}}
       {\left(r^{2}_{\rm H}+a^{2}e^{4\hat{\nu}_{\rm H}(1)}
        \right)^{2}}\,
  e^{2\hat{\nu}_{\rm H}(1)}
\end{equation}
at the axis. This can become negative for rapid rotation, namely
for
\begin{equation}  \label{alim}
  a>2M\frac{\sqrt{e^{4\hat{\nu}_{\rm H}(1)}+
                  \frac{2}{1-4\hat{\nu}_{{\rm H},y}(1)}}}
           {1+e^{4\hat{\nu}_{\rm H}(1)}+
            \frac{2}{1-4\hat{\nu}_{{\rm H},y}(1)}} \, .
\end{equation}
The static limit is located where
\begin{equation}  \label{staticlimit}
  \sqrt{\Delta}{\cal P}={\cal R}\sin\theta.
\end{equation}

Writing the metric in the Weyl-Lewis-Papapetrou form (\ref{WLP}),
the asymptotics of the gravitational potential $\nu$ and of the
dragging angular velocity $\omega$,
\begin{equation}  \label{nu,omega,asymp}
  \nu=-\frac{M+\hat{M}}{r}+O(r^{-2}), \;\;\;\;\;
  \omega=\frac{2a(M+2\hat{M})}{r^{3}}+O(r^{-4}),
\end{equation}
tell us that $M+\hat{M}$ is the total mass and $(M+2\hat{M})a$ is
the total angular momentum of the solution; $\hat{M}$ denotes the
mass of the seed spacetime for which we assumed the asymptotics
$\hat{\nu}=-\hat{M}/r+O(r^{-2})$.

In the static case ($a=0$), the result represents a superposition
of a given seed with a Schwarzschild black hole, with $u$ playing
the role of an interaction term ---
$\nu=\hat{\nu}+\nu_{\rm Schw}$,
$\lambda=\hat{\lambda}+\lambda_{\rm Schw}-u$.
Thus one can take just the ``external source'' as a seed, the
black hole is ``supplied'' by the soliton method itself.
Hopefully, this applies to a general, stationary case as well.
Actually, our main aim has been to learn whether the
Belinskii-Zakharov technique could yield the field of a rotating
black hole surrounded by an axisymmetric disc.

It seems that for small enough (though by far not negligible)
values of $a$ the solution (\ref{metric})--(\ref{uv2}) does not
contain singularities on or above the horizon. However, it is
quite extensive to calculate and analyse curvature invariants,
even if using computer algebra. Coordinates were found in which
metric is regular on the horizon (they have the meaning of the
Kerr ingoing/outgoing coordinates there), but this is not enough
to claim the absence of a singularity. Unfortunately, we found
the horizon to be pinched rather than smooth in the equatorial
plane. Indeed, the proper azimuthal circumference
$2\pi(g_{\phi\phi})_{\rm H}$ of the horizon has a sharp minimum
in the equatorial plane if $a>0$ (and $\hat{M}>0$).
It is to be clarified whether this pathology is physical or the
Boyer-Lindquist--type coordinates just do not cover the whole
manifold. Most probably, however, there appears a supporting
surface between the hole and the external disc.

\subsection{Stationary superposition cultivated from the inverted
            first Morgan-Morgan disc}
\label{soliton}

Some formulas were already specified to the case of a thin
equatorial source in \cite{ZelS00}, in particular, those for the
calculation of the energy-momentum tensor and of integral
quantities characterizing the resulting spacetime.
Let us mention below several properties of the metric
(\ref{metric})--(\ref{uv2}) obtained for the inverted first
Morgan-Morgan disc \cite{Sem02b}, i.e. for $\hat{\nu}$ given by
(\ref{nudisc}) (with the original disc mass now denoted by
$\hat{M}$). We saw in section \ref{singul} that this disc is
singular at the rim. However, its other properties are quite
satisfactory and the singularity is only caused by a sharp
increase of density, so it is reasonable to study first this
simple case and only later to embark on superpositions generated
from ``less-singular'' seeds of inverted higher Morgan-Morgan
discs (section \ref{higher}).

On the horizon,
\begin{equation}  \label{nuH}
  \hat{\nu}=\hat{\nu}_{\rm H}=
 -\frac{2\hat{M}}{\pi k^{3}|y|^{3}}
  \left[(k^{2}y^{2}+b^2){\rm arctan}\frac{k|y|}{b}-bk|y|\right].
\end{equation}
This reduces to $\hat{\nu}_{\rm H}(y=0)=-\frac{4\hat{M}}{3\pi b}$
in the equatorial plane. The main horizon properties
(\ref{A},\ref{kappa,omegaH}) only depend on the axial value
$\hat{\nu}_{\rm H}(y=1)$.
In \cite{Sem02b} it is shown that with growing mass or decreasing
inner radius of the disc the horizon inflates, its surface
gravity weakens and its rotation slows down. The presence of the
external source weakens the dependence of the horizon curvature
on $a/M$.
The existence of the sharp equatorial contraction, revealed by
the behaviour of the proper horizon circumference, has been
supported by plotting the isometric embedding (\ref{embedd}).

The only undetermined quantities in the metric (\ref{metric}) are
the functions $u$ and $v$. They are fixed on the axis by the
boundary conditions
$u(y=\pm1)=0$, $v(y=\pm1)=\pm2\hat{\nu}(|y|=1)$.
On the horizon, one easily obtains
$u_{\rm H}(y)=2\hat{\nu}_{\rm H}(1)-2\hat{\nu}_{\rm H}(y)$,
$v_{\rm H}(y)=2\hat{\nu}_{\rm H}(1)\cdot{\rm sign}\,y=
 \mp{\rm const}$.
In order that the spacetime be reflectionally symmetric, the seed
potential $\hat{\nu}$ must be an even function of $y$. Then
$u$ is also even in $y$, whereas $v$ is odd. Details of the
courses of $u$ and $v$ are determined, according to the equations
(\ref{uv1}) and (\ref{uv2}), by $\hat{\nu}$. In our case,
$\hat{\nu}$ (\ref{nudisc}) is continuous and negative. It is
maximal on the axis, having $\hat{\nu}_{,r}>0$ and
$\hat{\nu}_{,\theta}=0$ there
[while
 $\hat{\nu}_{,y}(y=\pm1)
 \stackrel{\scriptscriptstyle >}{\scriptscriptstyle <}0$].
In the equatorial plane $\hat{\nu}$ is minimal;
$\hat{\nu}_{,r}$ is continuous there, being negative/positive
below/above the radius
$\rho=\sqrt{3/2}\,b$ ($\Leftrightarrow r=M+\sqrt{k^{2}+b^{2}}$)
where $\hat{\nu}$ assumes its global minimum of
$-(2/3)^{3/2}(\hat{M}/b)$. The latitudinal gradient
$\hat{\nu}_{,\theta}$ is only zero below the inner radius of the
disc, while it has a finite jump across the disc, given by
(\ref{limit}):
\begin{equation}  \label{nu,theta,equat}
  \lim_{y\rightarrow 0^{\pm}}\hat{\nu}_{,\theta}=
 -(r-M)\lim_{z\rightarrow 0^{\pm}}\hat{\nu}_{,z}=
  \mp\frac{4\hat{M}b}{\pi\Delta^{2}}(r-M)\sqrt{\Delta-b^{2}} \,.
\end{equation}
At infinity, $\hat{\nu}=-\hat{M}/r+O(r^{-2})$ which implies the
asymptotics
\begin{equation}  \label{uv,asymp}
  u=\frac{\hat{M}k\sin^{2}\theta}{r^{2}}+O(r^{-3}), \;\;\;\;\;\;
  v=-\frac{2\hat{M}y}{r}+O(r^{-2}).
\end{equation}

The values of $u$ and $v$ at general locations must be computed
numerically, one can only proceed further analytically in the
equatorial plane. There, equations (\ref{uv1}) and (\ref{uv2})
yield latitudinal derivatives
\begin{equation}
  u_{,\theta}=-\frac{2k}{r-M}\hat{\nu}_{,\theta}, \;\;\;\;\;\;
  v_{,\theta}=\frac{2\Delta}{r-M}\hat{\nu}_{,r}.
\end{equation}
Hence, $u$ increases in latitudinal direction from zero on the
axis to a maximum in the equatorial plane; this maximum is
smooth/sharp ($u_{,\theta}$ is zero / has a finite jump there)
below/above the inner disc radius. The global maximum of $u$ lies
where $\hat{\nu}_{,r}=0$ in the equatorial plane (i.e. at
$\rho=\sqrt{3/2}\,b$; $\hat{\nu}$ has a global minimum there).
The function $v$ is negative/positive in the upper/lower
hemisphere, having a finite jump across the equatorial plane;
this jump is constant --- equal to $-4\hat{\nu}_{\rm H}(y=1)$ ---
from the horizon up to the inner disc edge, then it decreases
with increasing radius. $v_{,\theta}$ is smooth across the
equatorial plane, being positive/negative above/below
$\rho=\sqrt{3/2}\,b$.

The radial derivatives of the equations (\ref{uv1}) and
(\ref{uv2}) reduce to
\begin{equation}
  u_{,r}=-\frac{2k}{r-M}\hat{\nu}_{,r}, \;\;\;\;\;\;
  v_{,r}=-\frac{2}{r-M}\hat{\nu}_{,\theta}
\end{equation}
in the equatorial plane. Substituting from (\ref{nu,above}),
(\ref{nu,below}) and (\ref{nu,theta,equat}), one comes to
expressions which can be integrated to explicit formulae for
$u(z=0,\rho<b)$, $u(z=0,\rho>b)$,
$v(z=0^{\pm},\rho<b)=\pm2\hat{\nu}_{\rm H}(y=1)$,
$v(z=0^{\pm},\rho>b)$ (see \cite{Sem02b}).
While $u(z=0)$ increases from the horizon to a maximum at
$\rho(=\sqrt{\Delta})=\sqrt{3/2}\,b$ and then decreases to zero
at radial infinity, $v(z=0^{\pm})$ increases/decreases to zero
monotonically from $\rho=b$.

Using the equatorial expressions for $u$ and $v$, one can, for
example, graph the dependence of the equatorial radius of the
static limit on $\hat{M}$, $b$ and $a$. We have found that the
ergosphere widens both with $a/M$ and with $\hat{M}$. One can
also plot the radial course of the light-like, dragging and
geodesic angular velocities
$\Omega_{\stackrel{\rm max}{\scriptscriptstyle\rm min}}$,
$\omega$ and $\Omega_{\pm}$.

\subsection{Extreme limit}
\label{extreme}

We assumed an under-extreme case ($k>0$) in \cite{ZelS00} in
deriving the general metric (\ref{metric})--(\ref{uv2}). However,
it can be checked where the quantities go in the extreme
{\em limit} of $k\rightarrow 0$, i.e. $a\rightarrow M$. One
finds that they behave as in the extreme limit of the pure Kerr
solution.

The event horizon approaches $r_{\rm H}=M$, its area, surface
gravity and angular velocity reach the values
\begin{equation}  \label{A,kappa,omegaH}
  A=8\pi M^{2}\cosh2\hat{\nu}_{\rm H},
  \;\;\;\;\;\;
  \kappa=0, \;\;\;\;\;\;
  \omega_{\rm H}=\frac{1}{2M},
\end{equation}
where $\hat{\nu}_{\rm H}=-\frac{4\hat{M}}{3\pi b}$ for the
inverted first Morgan-Morgan disc.
The Gaussian curvature of the horizon works out at
\begin{equation}
  C_{\rm H}=
 -\frac{4\pi e^{2\hat{\nu}_{\rm H}}\cosh^{3}2\hat{\nu}_{\rm H}}
       {A}
  \frac{(1+|y|)^{3}e^{4\hat{\nu}_{\rm H}}+
        (1-|y|)^{3}e^{-4\hat{\nu}_{\rm H}}-
        6\sin^{2}\theta}
       {\left[1+\left(|y|\,\cosh2\hat{\nu}_{\rm H}+
              \sinh2\hat{\nu}_{\rm H}\right)^{2}\right]^{3}}
\end{equation}
and reduces to
\begin{equation}
  C_{\rm H}(y=1)=
 -\frac{1}{2M^{2}\cosh2\hat{\nu}_{\rm H}}=
 -\frac{4\pi}{A},
\end{equation}
\begin{equation}
  C_{\rm H}(y=0)=
  \frac{3-\cosh4\hat{\nu}_{\rm H}}
       {M^{2}\cosh^{4}2\hat{\nu}_{\rm H}}
  e^{2\hat{\nu}_{\rm H}}
\end{equation}
at the axis and in the equatorial plane.

In the extreme limit, $u$ is seen to vanish. The equatorial
behaviour of $u$ and $v$ is important in discussion of the
singularity location $\Sigma=0$.
In \cite{ZelS00} it was shown that this singularity cannot lie
above the horizon (at least) if $a/M$ is sufficiently small. For
our seed (\ref{nudisc}), however, $\Sigma(r>r_{\rm H})$ is
positive in general even in the extreme limit. Actually, the
first term of (\ref{Sigma}) then goes over to
$[r-M(r-M)u/k]^{2}$. Using the expression for the equatorial
maximum of $u(k\rightarrow 0)$, the root $r$ of the above term
must satisfy
\begin{equation}
  \frac{r}{M}\left(\frac{\hat{M}}{M}-3\frac{b^{2}}{M^{2}}\right)
  \geq\frac{\hat{M}}{M}.
\end{equation}
This can only hold (for some positive $r$) if
$\hat{M}/M>3b^{2}/M^{2}$.
Such a situation can be considered in principle, but it is, in
fact, excluded astrophysically: $\hat{M}/M$ is typically claimed
to be much less than 1, while a realistic value of $3b^{2}/M^{2}$
amounts to the order of ten.

The static-limit surface $r_{0}(\theta)$ must be found
numerically as equation (\ref{staticlimit}) remains rather
cumbersome even in the extreme limit. Regarding the limiting
forms of $u$ and $v$, one can, however, infer the effect of the
external disc on the equatorial value of $r_{0}$. It is given by
\begin{equation}
  (r-M)(1-Mu/k)=M\cosh v, \;\;\;\;\;\; {\rm thus} \;\;\;\;\;\;
  r-M>M
\end{equation}
in the extreme limit, which means that $r_{0}(y=0)$ is greater
than its Kerr value ($2M$).

It is quite difficult to express, in terms of the parameters
$\hat{M}$, $b$, $M$ and $a$, the contributions of the black hole
and of the disc to the total mass and angular momentum of the
spacetime. However, the crucial Komar formula for the angular
momentum of the black hole $J_{\rm H}$, treated in equations
(154)--(156) of \cite{ZelS00}, can at least be integrated in the
extreme-limit case, to end up with just
\begin{equation}
  J_{\rm H}=M^{2}, \;\;\;\;\;
  M_{\rm H}=M, \;\;\;\;\;
  {\cal M}=\hat{M}, \;\;\;\;\;
  {\cal J}=2M\hat{M}.
\end{equation}

Another interesting feature of the $k\rightarrow 0$ limit of our
superposition is the vanishing of the ``external'' gravitational
field $\hat{\nu}_{,r}$, $\hat{\lambda}_{,r}$ on the horizon
\cite{Sem02a}. This effect of expulsion of the external
(stationary axisymmetric) fields from rotating (and/or charged)
black holes, analogous to the Meissner effect in
(super)conductors, was observed in magnetic fields before (see
the recent survey \cite{BicL00}).

\section{Concluding remarks}
\label{concluding}

In sections \ref{static} and \ref{stationary}, the classes of
static/stationary axisymmetric superpositions were presented for
a black hole with an external thin equatorial disc. They give
examples of the effects that can occur if the self-gravity of
the disc is not completely negligible. The sequence of stable
static discs with minimal possible radii, generated in section
\ref{sequence}, does {\em not} (necessarily) describe the
behaviour of a real accretion disc with its mass increasing (or
its inner radius changing). Actually, static annular discs need
not be of the given (inverted first Morgan-Morgan) type (and need
not remain within this class when parameters are changed
dynamically). Also, real accretion discs should be stationary
rather than static. The results reviewed in the previous
sections, however, allow for various generalizations. First,
instead of the first Morgan-Morgan solution, singular at the rim,
one can superpose an inversion of {\em any} member of the
Morgan-Morgan family of static counter-rotating\footnote
{However artificial the counter-rotating interpretation may seem,
 counter-rotating stellar discs in galaxies have become an
 observational matter \cite{TreY00}.}
finite thin discs (\ref{nu(m)inv}). Second, a different annular
disc can be employed, not belonging to the Morgan-Morgan family
at all (e.g. the inverted isochrone disc of Klein \cite{Kle97})
--- and possibly even a thick one (here we only mean the
exterior, vacuum field). The stationary metric
(\ref{metric})--(\ref{uv2}) applies to {\em any} vacuum static
axisymmetric seed, too.

We will try to clarify the nature of the pathology which occurs
in the equatorial plane of the metric
(\ref{metric})--(\ref{uv2}).
If it corresponds to a thin layer of mass rather than to a real
singularity, it might only be a consequence of the fact that
the Boyer-Lindquist coordinates cut off a certain part of the
manifold. It is more probable, however, that this layer
represents a supporting surface which bears witness that the
system's stationarity is only artificial.
What remains an open question in any case is the structure of the
black-hole interior, altered by the presence of the external
source. We hope to be able to answer it in future.

On a general level, one could conclude with the interview with
Profs. W. B. Bonnor and R. Penrose which Ji\v{r}\'{\i}
Bi\v{c}\'ak organized in 1968,\footnote
{It was exactly in 1968 that Prof. J. A. Wheeler coined the word
 ``black hole'', but he was only interviewed by J. B. several
 years later \cite{Bic78}.}
about mathematics, relativity and ``brain drain'' \cite{Bic69}
(it was a pretty hot topic at the time when ``brain strain'' was
to continue in Czechoslovakia, as Prof. Kucha\v{r} could
confirm). In his answer to the question of in which direction to
expect progress in relativity, Prof. Bonnor remarked: ``It is
also possible that quasars will indeed have something to do with
gravitational collapse.'' As I noticed in the literature, at
those days this was not yet a theme for a gentleman to speak
about in a decent astronomical society. Although the
``collapsed-matter'' interpretation of the quasi-stellar radio
sources \cite{Sal64} and the ``collapsed-star'' interpretation of
X-ray binaries \cite{ZelG66} were both introduced by ``ApJ''.
(The very importance of accretion as an energy source was
apparently first recognized by \cite{Zel64}.)

Nowadays, ApJ offers ``Viewing the shadow of the black hole at
the Galactic center'' within the next few years \cite{FalMA00}.
Nevertheless, will not a ({\em gravitational}) ``{\em shine} of a
black hole'' be delivered sooner? Gravitational radiation may
prove medicinal for rheumatism or elsewhere. Will travel agencies
cite papers like \cite{Teo98,Kra00} when offering flights
intended for ``gravitational basking''?
In any case, the abundance of event horizons in the last years'
NASA {\tt ADS} and {\tt astro-ph} libraries supports Prof.
Bonnor's tip. Today it is in fact impossible to follow, at
least in some detail, but a narrow sector of the black hole
research. Not within relativity, but mainly not within
astrophysics. In the bibliography below, we try to recommend
those papers which should be interesting both for physicists and
astrophysicists, mostly avoiding pure observational and pure
mathematical references.

By far not all the aspects of (``classical'' general
relativistic) black holes were mentioned in the survey part of
this paper. The very question of their origin was only touched
in connection with the cosmic censorship, with the formation of
heavy discs and with the decay of perturbations. Not a single
word about thermodynamics and quantum effects, about cosmology
and dark matter, about primordial black holes, about economical
and ecological aspects, about Tunguska impact \dots

Also, all the black holes considered in the present paper are
spheroidal. Recently, {\em toroidal} holes have been investigated
as another intriguing possibility, not only by theoreticians but
also as a (rather non-orthodox) candidate for active galactic
nuclei \cite{Spi00,PomHR00}.
Toroidal black holes can only exist provided that some of the
Hawking \& Ellis' assumptions are relaxed. Gannon \cite{Gan76}
proved the possibility of a smooth toroidal horizon by dropping
stationarity; the period for which a toroidal black hole can
persist is, however, very limited: the hole in the torus must
close up before a light ray can pass through \cite{JacV95}.
Toroidal horizons were really observed as a temporary stage of a
numerical gravitational collapse (\cite{Teu98} and references
therein). The case of a toroidal horizon with a (ring)
singularity outside (thus violating the predictability
assumption) was obtained in \cite{Pet79}. Other possibility is to
relax the asymptotic flatness and allow for a non-zero
cosmological constant $\Lambda$. A metric involving (long-term
stable) higher-genus horizons in an asymptotically
anti--de Sitter spacetime is quite simple; it was already
obtained as a result of a gravitational collapse and generalized
to charged and rotating cases (see \cite{KleMV98} and references
therein). Astronomical observations indicate that $\Lambda>0$,
but it is not excluded that the stabilizing attractive effect
similar to that of $\Lambda<0$ could be provided by a host galaxy
or by a massive accretion disc around the hole (!).
(For many more hints, astrophysical connotations and further
references, see \cite{Spi00}.)

\vspace{3mm}

In connection with the Weber's putative detection of
gravitational waves in 1971, R. Penrose speculated about a naked
singularity in the Galactic centre \cite{Pen72}, but at the end
he stated:
``This would remove the observational discrepancies, although it
would be a radical explanation\dots In comparison black holes can
now be regarded as being `conventional'; indeed, for this reason
they must be preferred a priori in any attempt at explanation.''
Our essay {\em has been} a priori 'conventional', preferring
black holes, and of spheroidal topology.
Well, Penrose then continued:
``Yet nature does not always prefer conventional explanations,
least of all in astronomy.''

\section*{Acknowledgements}

I thank my former Ph.D. students M. \v{Z}\'a\v{c}ek and T.
Zellerin, the results reviewed in sections \ref{static} and
\ref{stationary} were obtained in collaboration with them.
We were encouraged by J. Bi\v{c}\'ak, V. Karas, P. Krtou\v{s}
and T. Ledvinka.
Support from the Czech grants GACR-202/99/0261, GACR-205/00/1685
and GAUK-141/2000 is acknowledged.

\end{document}